\documentclass[aip,reprint,superscriptaddress,amsmath,amssymb]{revtex4-2}
\usepackage[utf8]{inputenc}
\usepackage{graphicx}
\usepackage{physics}
\usepackage{hyperref}
\usepackage{booktabs}
\usepackage{xcolor}
\usepackage{textcomp}
\begin{document}

% Use the \preprint command to place your local institutional report
% number in the upper righthand corner of the title page in preprint mode.
% Multiple \preprint commands are allowed.
% Use the 'preprintnumbers' class option to override journal defaults
% to display numbers if necessary
%\preprint{}

% repeat the \author .. \affiliation  etc. as needed
% \email, \thanks, \homepage, \altaffiliation all apply to the current
% author. Explanatory text should go in the []'s, actual e-mail
% address or url should go in the {}'s for \email and \homepage.
% Please use the appropriate macro foreach each type of information

% \affiliation command applies to all authors since the last
% \affiliation command. The \affiliation command should follow the
% other information
% \affiliation can be followed by \email, \homepage, \thanks as well.

% previous title:
% title{Hysteresis Measurements of x-cut MgO-doped Thin-Film Lithium Niobate: Understanding High Coercive Fields in Periodic Poling Applications}
\title{Ferroelectric Switching in x-cut MgO-doped Thin-Film Lithium Niobate: Implications for Periodic Poling and Domain-Wall Electronics}

\author{Niels Staal}
\affiliation{Department of Physics, Paderborn University, Warburger Straße 100, 33098 Paderborn, Germany}
\affiliation{Institute for Photonic Quantum Systems (PhoQS), Paderborn University, Warburger Straße 100, 33098 Paderborn, Germany}

\author{Michael~Rüsing}
\email{michael.ruesing@uni-paderborn.de} % VERIFY address
\affiliation{Department of Physics, Paderborn University, Warburger Straße 100, 33098 Paderborn, Germany}
\affiliation{Institute for Photonic Quantum Systems (PhoQS), Paderborn University, Warburger Straße 100, 33098 Paderborn, Germany}

\author{Tobias Babai-Hemati}
\affiliation{Department of Physics, Paderborn University, Warburger Straße 100, 33098 Paderborn, Germany}
\affiliation{Institute for Photonic Quantum Systems (PhoQS), Paderborn University, Warburger Straße 100, 33098 Paderborn, Germany}

\author{Boris Koppitz}
\affiliation{Institute of Applied Physics, Technische Universität Dresden, Nöthnitzer Straße 61, 01187 Dresden, Germany}

\author{Nivedita Vishnukumar}
\affiliation{Department of Physics, Paderborn University, Warburger Straße 100, 33098 Paderborn, Germany}
\affiliation{Institute for Photonic Quantum Systems (PhoQS), Paderborn University, Warburger Straße 100, 33098 Paderborn, Germany}

\author{Laura Bollmers}
\affiliation{Department of Physics, Paderborn University, Warburger Straße 100, 33098 Paderborn, Germany}
\affiliation{Institute for Photonic Quantum Systems (PhoQS), Paderborn University, Warburger Straße 100, 33098 Paderborn, Germany}

\author{Noah Spiegelberg}
\affiliation{Department of Physics, Paderborn University, Warburger Straße 100, 33098 Paderborn, Germany}

\author{Philipp Mues}
\affiliation{Institute for Photonic Quantum Systems (PhoQS), Paderborn University, Warburger Straße 100, 33098 Paderborn, Germany}

\author{Sebastian Lengeling}
\affiliation{Department of Physics, Paderborn University, Warburger Straße 100, 33098 Paderborn, Germany}
\affiliation{Institute for Photonic Quantum Systems (PhoQS), Paderborn University, Warburger Straße 100, 33098 Paderborn, Germany}

\author{Tobias Schwabe}
\affiliation{Institute for Photonic Quantum Systems (PhoQS), Paderborn University, Warburger Straße 100, 33098 Paderborn, Germany}
\affiliation{Heinz Nixdorf Institute, Department of Electrical Engineering, Paderborn University, Warburger Straße 100, 33098 Paderborn, Germany}

\author{J. Christoph Scheytt}
\affiliation{Heinz Nixdorf Institute, Department of Electrical Engineering, Paderborn University, Warburger Straße 100, 33098 Paderborn, Germany}
\affiliation{Institute for Photonic Quantum Systems (PhoQS), Paderborn University, Warburger Straße 100, 33098 Paderborn, Germany}

\author{Christof Eigner}
\affiliation{Institute for Photonic Quantum Systems (PhoQS), Paderborn University, Warburger Straße 100, 33098 Paderborn, Germany}

\author{Lukas M. Eng}
\affiliation{Institute of Applied Physics, Technische Universität Dresden, Nöthnitzer Straße 61, 01187 Dresden, Germany}
\affiliation{ctd.qmat: Dresden-Würzburg Cluster of Excellence—EXC 2147, Technische Universität Dresden, 01062 Dresden, Germany}

\author{Laura~Padberg}
\affiliation{Department of Physics, Paderborn University, Warburger Straße 100, 33098 Paderborn, Germany}
\affiliation{Institute for Photonic Quantum Systems (PhoQS), Paderborn University, Warburger Straße 100, 33098 Paderborn, Germany}

\author{Christine~Silberhorn}
\affiliation{Department of Physics, Paderborn University, Warburger Straße 100, 33098 Paderborn, Germany}
\affiliation{Institute for Photonic Quantum Systems (PhoQS), Paderborn University, Warburger Straße 100, 33098 Paderborn, Germany}

\date{\today}

\begin{abstract}
Thin-film lithium niobate (TFLN) revives a mature ferroelectric as a wafer-scale platform for integrated photonics and, increasingly, for domain-wall electronics such as nonvolatile memories and memristors. Both applications rely on controlled ferroelectric switching, which is poorly understood in TFLN: periodic poling consistently requires electric fields above 30~kV/mm, far above the coercive fields of about 21~kV/mm and 4.5~kV/mm of bulk congruent and 5~mol\% MgO-doped lithium niobate, respectively. Here, we present systematic ferroelectric hysteresis measurements on x-cut, 5~mol\% MgO-doped TFLN with coplanar electrodes as used for periodic poling, combining the positive-up-negative-down (PUND) method with second-harmonic microscopy (SHM). We obtain a spontaneous polarization of $(0.72\pm0.02)$~C/m$^2$, consistent with bulk values, while the coercive field varies between 24 and 40~kV/mm with the voltage ramp rate and depends on the electrode geometry. We identify three contributions to the high coercive fields: (i) fast voltage ramps, (ii) finger electrodes, and (iii) the coplanar surface-electrode geometry. Together with bulk data, the coercive field follows a single power law of the ramp rate over five orders of magnitude, showing that the high fields are dominated by switching dynamics rather than by intrinsic thin-film properties. Moreover, the apparent internal field depends on the electrode geometry, and reverse switching leaves buried, electrically isolated domains. These results provide guidelines for scalable, fab-compatible periodic poling---one of the principal bottlenecks of TFLN device integration---and for the write and erase operation of TFLN domain-wall devices.
\end{abstract}

\maketitle

\section{Introduction}
Lithium niobate (LN) and lithium tantalate (LT) are among the most established ferroelectric single crystals. They are grown as large single crystals for more than six decades and are used on an industrial scale, e.g., in surface-acoustic-wave filters for mobile communication or for electro-optic and nonlinear optical devices \cite{Ballman1965,Weis1985,Hashimoto2000}. With the commercial availability of single-crystalline thin films on insulator on wafer scale over the last decade, LN and LT have seen a remarkable revival. Thin-film lithium niobate (TFLN) has revolutionized integrated photonics by combining strong ferroelectric properties with large electro-optic and nonlinear-optical coefficients, a high refractive-index contrast, and commercial wafer-scale availability \cite{zhu2021,Vazimali2022,Boes2023,Labbe2025,Hu2025}. This enables applications ranging from high-speed optical modulation for data communication \cite{Xue2026,Li2026}, terahertz and microwave-to-optical conversion \cite{Juneghani2024,Shao2019}, laser-frequency conversion \cite{Lohmann2025,Lu2019}, and frequency-comb generation \cite{Wang2025} to numerous quantum-optical applications in quantum communication, quantum simulation, or quantum computing \cite{Sund2023,Bernardi2026,Babel2023}. At the same time, TFLN is compatible with hybrid and heterogeneous integration with silicon photonics, silicon nitride, and silicon electronics \cite{Zhao2019,He2019,Churaev2023}, and it enables new acoustic devices at improved frequencies and decreased footprint \cite{Plessky2019}.

A key feature that distinguishes LN from many other photonic and electronic materials, like silicon or many III-V systems, is its ferroelectricity: the direction of the spontaneous polarization can be switched locally and nonvolatilely by an electric field. This domain engineering provides electric, reconfigurable control over all properties that depend on the sign of the polarization, such as the second-order nonlinear and electro-optic coefficients, the piezoelectric response, and surface and wall charges \cite{Gopalan2007,Boes2023}. In TFLN, two major fields of application exploit this capability: domain-wall (DW) electronics and periodic poling for nonlinear optics.

Owing to its wide band gap, single-domain LN is highly insulating, whereas charged DWs can be conductive by many orders of magnitude \cite{Schroeder2012,Godau2017,Werner2017,Beccard2022,Beccard2023}. Because DWs can be written, moved, modified, and erased by voltage pulses, they act as reconfigurable, nanoscale conducting channels \cite{Catalan2012,MeierSelbach2022,Sharma2022}. In LN thin films, this has enabled nonvolatile DW memories integrated on Si with high on/off ratios and endurance \cite{Jiang2020,Jiang2024roadmap}, DW memristors with many conductance states and synaptic plasticity for neuromorphic computing \cite{McConville2020,Chaudhary2020}, tunable DW memories \cite{Kaempfe2020}, and multilevel DW devices for in-memory computing \cite{EPL2025}. Many of these concepts use in-plane domains written by coplanar electrodes on x- or y-cut films due to the ease of structuring by 2D lithography \cite{Jiang2020}. In all of them, the write voltage and speed, the retention, and the reliability of the erase operation are determined by the switching process, i.e., by the coercive field and its dependence on the pulse parameters, by internal fields (imprint), defects, and by the completeness of reverse switching. Here, poling needs to be understood locally to determine performance and lifetimes of DW-based devices. As for emergent ferroelectrics, such as doped HfO$_2$ or AlScN \cite{Boescke2011,Fichtner2019,Mikolajick2021}, the quantitative understanding and control of these parameters remain key challenges. These might be well characterized for bulk LN, but remain only loosely explored for TFLN and in-plane switching with surface electrodes.

In nonlinear optics, periodic poling creates quasi-phase-matching (QPM) structures for efficient frequency conversion by compensating the phase mismatch between the interacting waves \cite{byer1997,Zhao2019,Hu2025,Labbe2025}. Here, the challenge is of a different kind: domain gratings with periods of a few micrometers have to be inverted homogeneously over lengths of millimeters to centimeters, typically with short, high-field pulses far from quasi-static switching conditions, and the device performance depends directly on the shape, homogeneity and duty cycle of the domains on this large scale. Therefore, the fabrication, scaling, and understanding of periodic poling are key enabling technologies in TFLN integrated photonics \cite{wang2019,saravi2021,mackwitz2016,bocchini2025}.

\begin{table*}[htpb]
\centering
\caption{Overview of reported conditions for periodic poling of bulk z-cut lithium niobate (LN) and x-cut thin-film lithium niobate (TFLN) of congruent and MgO-doped composition. Because domain inversion during periodic poling is a highly dynamic and complex process, the applied fields are usually chosen larger than the coercive field. The listed fields therefore represent recipes for successful periodic poling rather than the coercive field of the material. The peak field is estimated as peak voltage divided by electrode spacing. Please note, some works used more complex poling pulses, which this table cannot fully reflect. It is rather meant as a approximate guideline to the most critical parameters.}
\label{tab:ln_coercive_fields}
\small
\begin{tabular}{p{5.0cm} c c c c c r}
\toprule
\textbf{\shortstack{Material details and\\ poling conditions}} & \textbf{\shortstack{Electrode\\ spacing ($\mu$m)}} & \textbf{\shortstack{Peak\\ voltage (V)}} & \textbf{\shortstack{Ramp rate\\ (kV/(mm\,s))}} & \textbf{\shortstack{Approx. peak\\ field (kV/mm)}} & \textbf{\shortstack{Peak-voltage\\ duration (ms)}} & \textbf{Ref.} \\
\midrule
\multicolumn{7}{c}{\textbf{Z-cut bulk crystals}} \\
\midrule
congruent LN & $500$ & -- & -- & $\approx 21$ & $\approx 400$ & \cite{miller1998periodically} \\
\addlinespace[1pt]
congruent LN & -- & -- & -- & $\approx 24$ & $\approx 0.1$ & \cite{Yamada1993} \\
\addlinespace[1pt]
congruent LN & $500$ & $\approx 11250$ & -- & $\approx 23$ & $\approx 2$ & \cite{Pandiyan2008} \\
\addlinespace[1pt]
congruent LN & $500$ & $\approx 10375$ & $\approx 10$--$100$ & $\approx 20.75$ & $\approx 5$ & \cite{Batchko1999} \\
\addlinespace[1pt]
congruent LN & $500$ & $\approx 11500$ & -- & $\approx 23$ & $\approx 500$ & \cite{Peng1999} \\
\addlinespace[1pt]
congruent LN & $500$ & $\approx 11250$ & -- & $\approx 22$ & $\approx 1700$ & \cite{jin2005} \\
\addlinespace[5pt]
5~mol\% MgO:LN & $300$ & $\approx 2000$ & -- & $\approx 6.7$ & $\approx 4000$ & \cite{Nakamura2002} \\
\addlinespace[1pt]
5~mol\% MgO:LN & $2000$ & $\approx 10000$ & -- & $\approx 5$ & $\approx 1$ & \cite{Mizuuchi2004} \\
\addlinespace[1pt]
5~mol\% MgO:LN & $3000$ & $\approx 13500$ & -- & $\approx 4.5$ & $\approx 100000$ & \cite{Ishizuki2003} \\
\midrule
\multicolumn{7}{c}{\textbf{X-cut congruent (undoped) TFLN}} \\
\midrule
$t_{\text{TFLN}}=$300~nm, pre-etched waveguide & $8.5$ & $400$ & -- & $\approx 47.3$ & $\approx 5$ & \cite{Rao2019} \\
\addlinespace[1pt]
\textbf{y-cut}, $t_{\text{TFLN}}=$400~nm, electrodes etched into TFLN layer & $\leq 10$ & $\leq 400$ & -- & $\approx 40$ & $\approx 10$ & \cite{Rao2016} \\
\addlinespace[1pt]
$t_{\text{TFLN}}=$500~nm, multiple pulses & $\approx 9$ & $\approx 450$ & $\approx 10000$ & $\approx 50$ & $\approx 20$ & \cite{Chen2019} \\
\addlinespace[1pt]
$t_{\text{TFLN}}=$600~nm & $6$ & $\approx 240$ & -- & $\approx 40$ & $\geq 10$ & \cite{Niu2020} \\
\addlinespace[1pt]
$t_{\text{TFLN}}=$600~nm, pre-conditioning and nucleation pulses & $20$ & $\leq 1080$ & $\approx 38800$ & $\leq 54$ & $\approx 3$ & \cite{Younesi2021} \\
\addlinespace[1pt]
$t_{\text{TFLN}}=$700~nm & $12$ & $\approx 600$ & -- & $\approx 50$ & -- & \cite{mackwitz2016} \\
\addlinespace[1pt]
$t_{\text{TFLN}}=$800~nm, pre-pulses & $\approx 15$ & $\approx 450$ & $\approx 10000$ & $\approx 30$ & $\leq 5$ & \cite{Ayhan2025} \\
\addlinespace[1pt]
$t_{\text{TFLN}}=$800~nm, three long rectangular pulses & $15$ & $\approx 400$ & -- & $\approx 26.6$ & $\approx 500$ & \cite{Liu2026} \\
$t_{\text{TFLN}}=$500~nm, between 3 to 1000 rectangular pulses & $9.2$ & $\approx 280 -380$ & -- & $\approx 28.6-41.3$ & $\approx 50$ & \cite{Zhang2020} \\
\midrule
\multicolumn{7}{c}{\textbf{X-cut 5~mol\% MgO-doped TFLN (MgO:TFLN)}} \\
\midrule
$t_{\text{TFLN}}=$250~nm, SiO$_2$ layer under electrodes & $20$ & $\leq 500$ & $\geq 100000$ & $\approx 25$ & $\approx 0.3$ & \cite{Park2022} \\
\addlinespace[1pt]
$t_{\text{TFLN}}=$300~nm & $20$ & $\leq 800$ & $\leq 40000$ & $\leq 40$ & $\leq 2$ & \cite{Zhao2020,Zhao2020b,Zhao2020c,ruesing2019} \\
\addlinespace[1pt]
$t_{\text{TFLN}}=$300~nm, bonded to silicon photonics & $10$ & $\leq 600$ & $\leq 60000$ & $\leq 60$ & $\leq 2$ & \cite{Zhao2019} \\
\addlinespace[1pt]
$t_{\text{TFLN}}=$600~nm & $40$ & $\approx 1250$ & -- & $\approx 31.3$ & -- & \cite{Bollmers2025} \\
\addlinespace[5pt]
$t_{\text{TFLN}}=$700~nm, SiO$_2$ layer under electrodes & $15$ & $\approx 550$ & $\approx 36600$ & $\approx 36.6$ & $\approx 1$ & \cite{Nagy2019} \\
\addlinespace[1pt]
$t_{\text{TFLN}}=$700~nm, SiO$_2$ layer under electrodes, pre-conditioning pulses & $15$ & $\approx 500$ & $\approx 33300$ & $\approx 33.3$ & $\approx 1$ & \cite{Nagy2020} \\
\addlinespace[1pt]
$t_{\text{TFLN}}=$200~nm, SiO$_2$ layer under electrodes, pre-anneal at 200\textdegree C & $15$ & $\approx 475$ & $\approx 36600$ & $\approx 31.7$ & $\approx 100$ & \cite{Doshi2024} \\
\addlinespace[5pt]
$t_{\text{TFLN}}=$300~nm, poling after etching, sidewall electrodes & $2$ & $120$ & $\leq 60000$ & $60$ & $\approx 1$ & \cite{Sabatti2025} \\
\addlinespace[1pt]
$t_{\text{TFLN}}=$300~nm, poling after etching, sidewall electrodes & $1$ & $80$ & $\leq 60000$ & $80$ & $\approx 1$ & \cite{Sabatti2025} \\
\addlinespace[1pt]
$t_{\text{TFLN}}=$600~nm, poling after waveguide etching & $15$ & $\approx 600$ & -- & $\approx 40$ & $\approx 1$ & \cite{Xin2025} \\
\addlinespace[1pt]
$t_{\text{TFLN}}=$600~nm, SiO$_2$ layer under electrodes, poling after waveguide etching, sidewall electrodes & $\leq 1.5$ & $\approx 127$ & -- & $\approx 85$ & $\approx 1$ & \cite{Franken2026} \\
\addlinespace[1pt]
$t_{\text{TFLN}}=$600~nm, poling of ridge waveguides without cladding & $20$ & $\leq 500$ & $\approx 136000$ & $\approx 25$ & $\approx 0.1$ & \cite{Yamen2026} \\
\addlinespace[1pt]
$t_{\text{TFLN}}=$600~nm, poling of ridge waveguides through 300~nm SiO$_2$ cladding at 250--300\textdegree C & $20$ & $\leq 800$ & $\approx 133000$ & $\approx 40$ & $\approx 0.1$ & \cite{Yamen2026} \\
\bottomrule
\end{tabular}
\\[2pt]
\end{table*}

Highly efficient QPM nonlinear-optical frequency converters are regularly reported, showing that poling processes have been developed successfully by numerous groups \cite{Zhao2020b,Babel:25,Wang2018,Xin2025,Hu2025,Lohmann2025,Lu2019,Sabatti2025,Jankowski2020}. Nevertheless, the fabrication of periodically poled TFLN (PPTFLN) remains challenging, e.g., regarding the reproducibility and homogeneity of the domains or the realization of small poling periods. Hence, still numerous works deal with various improvements ranging from (pre-)annealing over pre-pulses to ultra-small poling periods by sidewall poling \cite{Doshi2024,Nagy2019,Sabatti2025,Zhao2020b}. Moreover, the poling process itself remains poorly understood from a solid-state-physics perspective, which hinders systematic, theory-guided optimization.

The most common technique for ferroelectric domain inversion is electric-field poling via lithographically structured electrodes \cite{Chakkoria2025,Bollmers2025}, which can be scaled with standard nanofabrication techniques \cite{Chakkoria2025,Bollmers2025,Chen2019,Xin2025,Chen2024}. During periodic poling, an electric field exceeding the coercive field of the crystal is applied via the periodic electrodes, which locally inverts the spontaneous polarization \cite{Chakkoria2025,Bollmers2025}. For bulk 5~mol\% MgO-doped and congruent lithium niobate (LN), the coercive fields are reported to be about 4.5~kV/mm and 21--22~kV/mm, respectively \cite{Gopalan2007,Ishizuki2003}. In bulk periodic poling, fields slightly above the coercive field are commonly applied, and the process is stopped before full inversion has taken place. The poling result can be monitored ex situ by domain-visualization techniques \cite{Gruverman2019,Soergel2011,Roeper2024,Reitzig2022,Rix2022,Hegarty2022,Hempel2024,Ratzenberger2024}, or in situ, e.g., by optical imaging \cite{Ratzenberger2024,Kiseleva2025,Missey2000} or by monitoring the poling current and stopping at a predefined charge \cite{Kumar2026}. Table~\ref{tab:ln_coercive_fields} summarizes typical poling conditions for bulk z-cut LN and x-cut TFLN of congruent and MgO-doped composition. Since these are poling recipes rather than coercive-field measurements, the listed fields are naturally larger than the coercive fields. Strikingly, however, while for bulk z-cut crystals fields only 1--2~kV/mm above the coercive field are reported, the fields applied to TFLN regularly exceed the bulk coercive field by 20~kV/mm or more. This is especially pronounced for MgO-doped TFLN, where the applied fields exceed the bulk coercive field by a factor of six to ten.

This discrepancy has led to speculation about thin-film-specific material properties, e.g., defect structures caused by lithium out-diffusion, leakage at the electrode--film interface, or domain pinning at the SiO$_2$/TFLN interface \cite{bocchini2025,Nagy2019,Vazimali2022,Chang2016,Chen2024}. All of these explanations are plausible, since lithium deficiency, the defect structure, leakage, and the metal--LN interface are known to affect poling and charge transfer, and they are also studied in related fields such as electro-optic modulators and ferroelectric memories \cite{Hurtado2025,Gopalan2007,Kim2002,Zahn2026}. However, a clear picture of the underlying mechanisms in TFLN is still missing. While the resulting periodic domain structures are frequently studied, the poling process itself is usually not investigated.

Beyond the substantially higher fields, a second striking difference emerges when comparing bulk and TFLN poling (Table~\ref{tab:ln_coercive_fields}): bulk poling typically uses pulses that are one to three orders of magnitude longer, together with often much slower field ramps. At the same time, the switched cross-sections, and hence the switched charges, differ by many orders of magnitude. In TFLN, the film is thinner than 1~$\mu$m and the electrodes are typically a few millimeters long, resulting in cross-sections on the order of $10^{-9}$~m$^2$, whereas bulk samples are poled over areas of up to several square centimeters (on the order of $10^{-4}$~m$^2$). Consequently, bulk poling is often limited by the current that the high-voltage source can supply: the voltage drops once switching sets in, which results in a partially self-regulating process \cite{miller1998periodically,Mizuuchi2004,Yamada1993,Pandiyan2008,Nakamura2002,Batchko1999}. This also explains the comparatively low peak fields reported for bulk poling. Furthermore, bulk poling is typically performed at fields close to the coercive field $E_C$, where the domain-wall velocity depends most steeply on the applied field, which allows the switching process to be controlled precisely via the applied voltage \cite{miller1998periodically,Yamada1993}.

Hysteresis measurements of the polarization as a function of the electric field are the gold standard for ferroelectric characterization since the early research on these materials \cite{scott2000,Sawyer1930,Corbellini2016}. They probe the switching dynamics in situ by measuring the transferred charge, rather than evaluating the domain structure after poling. In this way, extrinsic contributions, such as leakage currents or the influence of the electric circuit (e.g., its capacitance), can be separated from the actual switching. Evaluating the time and field dependencies further provides insight into the role of nucleation, domain-wall pinning, and defects\cite{Shao2016}. Nevertheless, systematic hysteresis studies of x- or y-cut TFLN are notably absent from the literature, despite the widespread use of PPTFLN devices \cite{zhu2021,Franken2026,Zhao2020c}, in particular for electrode geometries that match those used in periodic poling. To our knowledge, only two related studies exist. Nagy and Reano measured hysteresis loops on x-cut MgO:TFLN and found a reasonable spontaneous polarization close to the bulk value, but also coercive fields of 30--40~kV/mm \cite{Nagy2020} substantially larger than bulk values. Their loops show a pronounced slope and deviate from the ideal rectangular shape, which may indicate leakage or limitations of the measurement equipment; further parameters, such as the ramp rate or the electrode geometry, were not varied. In another work, 43-nm-thick y-cut TFLN, tightly sandwiched between electrodes, was switched up to millions of times for ferroelectric-memory applications \cite{Hurtado2025}. This geometry, film thickness and crystal cut, however, is not comparable to the structures used in integrated optics. The goal of the present work is, therefore, to close this gap by measuring hysteresis loops of TFLN under different conditions and to identify the underlying physical mechanisms.

This work has two aims. First, it is intended to be accessible to readers without a strong background in ferroelectrics: Section~\ref{sec:theory} serves as a tutorial on the principles of hysteresis measurements and may be skipped by experts. Second, we present the first comprehensive hysteresis measurements of x-cut MgO-doped TFLN. We developed a setup capable of detecting switching currents in the (sub-)microampere range at kilovolt voltages, which enables to use the the positive-up-negative-down (PUND) method to separate ferroelectric switching from capacitive and leakage currents. To verify the poling result and the poling depth, we use second-harmonic microscopy (SHM), which is sensitive to incomplete poling in depth \cite{ruesing2019}. This allows us to determine the switched cross-section and hence to estimate the spontaneous polarization accurately. Our systematic investigation shows that the high coercive fields observed in TFLN are not intrinsic material properties but mainly result from (i) the application of poling pulses up to three orders of magnitude faster than in bulk poling and, to a lesser extent, from (ii) the use of finger electrodes and (iii) the coplanar surface-electrode geometry, which both limit nucleation and charge exchange. Beyond periodic poling, the results are directly relevant to TFLN-based DW electronics, which uses similar coplanar electrode geometries, fast write pulses, and charged DWs.

\section{Tutorial: Measuring hysteresis loops}
\label{sec:theory}

\subsection{Ferroelectricity and polarization}

% FIGURE TODO (Fig. 1, for the student) -- see changelog, section "Figures":
%  - check resolution; (a) shade the crystal side faces slightly;
%  - (b) bottom panel: label axis "Charge Q" (accumulated charge), not "I charge";
%    current and charge should change only during the rising flank of the voltage, not stretch into the falling flank;
%  - delete panel e) (if still present).
\begin{figure*}[t]
\centering
\includegraphics[width=0.95\textwidth]{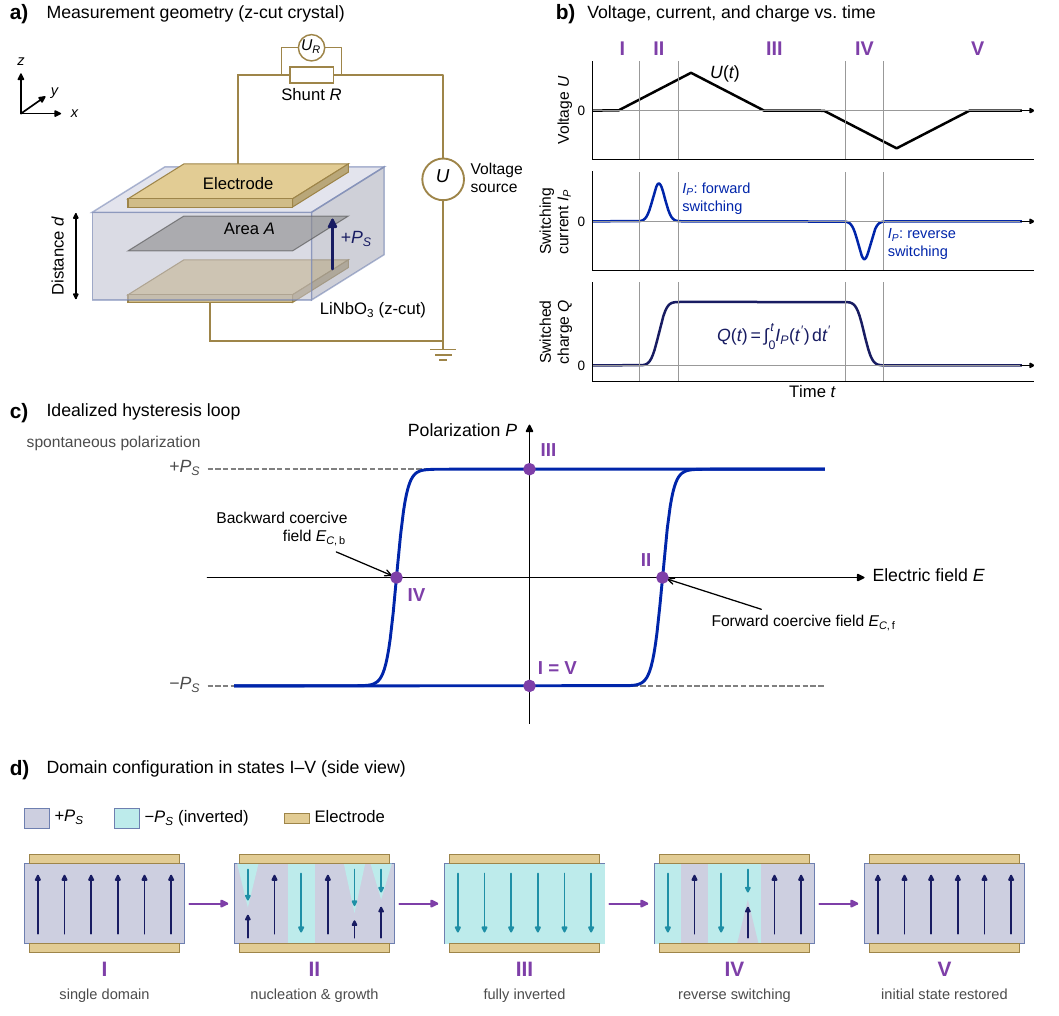}
\caption{Principle of a ferroelectric hysteresis measurement on a z-cut crystal. (a) Measurement geometry and circuit: a voltage $U$ is applied across a crystal of thickness $d$ via electrodes covering the area $A$; the current is obtained from the voltage drop $U_R$ across the shunt resistor $R$. The arrow marks the spontaneous polarization $+P_S$ of the as-grown crystal along $+z$. (b) Triangular voltage pulses of positive and negative polarity, the resulting switching current $I_P$, and the switched charge $Q(t)$; switching is completed before the voltage maximum. (c) Idealized polarization--field ($P$--$E$) hysteresis loop with spontaneous polarization $\pm P_S$ (dashed lines) and forward and backward coercive fields $E_{C,\mathrm{f}}$ and $E_{C,\mathrm{b}}$. $P$ and $E$ are taken along the direction of the applied field; therefore, the as-grown state appears at $P=-P_S$ (state I). (d) Side-view sketches of the domain configuration in the states I--V: nucleation of spike domains (II), complete inversion (III), and reverse switching (IV), which restores the initial state (V).}
\label{fig:hysteresis_states}
\end{figure*}

Ferroelectric materials possess a spontaneous polarization $\mathbf{P}_S$ that exists even in the absence of an external electric field. It arises from the ordered arrangement of electric dipoles in the crystal, i.e., from a built-in dipole moment in each unit cell, which requires a polar crystal structure without inversion symmetry \cite{sonin1974}. In lithium niobate, this dipole moment originates from the displacement of the Li$^+$ and Nb$^{5+}$ ions relative to the oxygen (O$^{2-}$) layers along the crystallographic $z$-axis \cite{Gopalan2007}. The lack of inversion symmetry also gives rise to the second-order nonlinear-optical and electro-optic properties, which underlines the close connection between ferroelectricity and the optical applications of LN. In an uniaxial ferroelectric, such as LN, the polarization is restricted to a single crystallographic axis (the $z$-axis), which results in only two stable states, $+\mathbf{P}_S$ and $-\mathbf{P}_S$ \cite{Gopalan2007}. Other ferroelectrics, such as barium titanate, which is currently being explored for integrated optics \cite{Moehl2025,Esfandiar2026}, are multiaxial and allow more orientations of the spontaneous polarization \cite{Kirbus2024}. This requires a careful consideration of the domain orientation relative to the poling electrodes and the crystal axes.

% FIGURE TODO (Fig. 2): add "0" to all voltage axes and the axis label "Voltage U" in panel (a);
%  make "+" and "=" signs less bold / slightly smaller; increase font size (at least as in Fig. 1);
%  yellow text is hard to read -> darker color; typo in panel f): "Negativ Down" -> "Negative Down".
\begin{figure*}[t]
\centering
\includegraphics[width=0.95\textwidth]{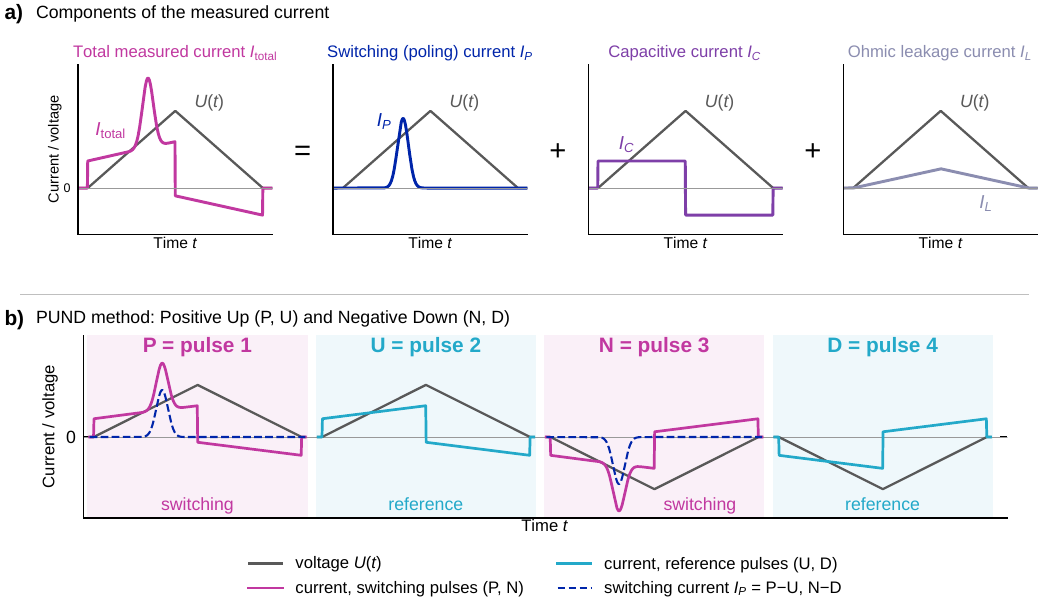}
\caption{(a) Components of the measured current during a triangular voltage pulse $U(t)$ (dark gray): the total current $I_\mathrm{total}$ is the sum of the switching current $I_P$, the capacitive current $I_C$, and the Ohmic leakage current $I_L$. (b) PUND sequence: the switching pulses P and N (magenta) invert the domains and contain all three components, whereas the reference pulses U and D (cyan), applied to the already switched state, contain only $I_C + I_L$. The switching current (blue, dashed) is obtained from the differences P$-$U and N$-$D.}
\label{fig:pund_method}
\end{figure*}

The macroscopic polarization $\mathbf{P}$ is related to the dipole moments $\mathbf{p}_j$ of the $N$ unit cells in a volume $V$ by
\begin{equation}
\label{eq:dipole}
\mathbf{P} = \frac{1}{V}\sum_{j=1}^{N} \mathbf{p}_j .
\end{equation}
Regions of uniform polarization are called ferroelectric domains, and the boundaries between them are called domain walls (DWs).

At a surface with area $A$ perpendicular to the polarization, the spontaneous polarization corresponds to a bound surface charge density $\sigma$:
\begin{equation}
\label{P_s}
P_S = \frac{Q}{A} = \sigma ,
\end{equation}

% COMMENT:  instead of -sigma (charge density)  it should be (-Q) (bound charge) The sentence still makes sense but if on the otherside if we have a different surface area then sigma will be different.

where $Q$ is the bound charge on this surface \cite{sonin1974}. The opposite surface carries the same bound charge with opposite sign $-\sigma$. In equilibrium, these bound charges are compensated by screening charges, e.g., free electrons or holes, charged defects, or adsorbates, which can also lead to chemical or structural changes in the topmost crystal layers depending on the polarization orientation \cite{Neumayer2018,Sanna2010}.

In a z-cut LN crystal, i.e., with the $z$-axis perpendicular to the surface, domains usually terminate at the crystal surfaces. In x-cut TFLN, in contrast, the $z$-axis lies in the plane of the film, and (periodically) poled domains terminate \emph{inside} the film, forming charged head-to-head and tail-to-tail DWs [see Fig.~\ref{fig:hysteresis_SHG}(e) below].

\subsection{Hysteresis and domain switching}

A ferroelectric is a polar material whose spontaneous polarization can be reversed by an external electric field $E$ without melting or structurally damaging the material. Note that not every polar material is ferroelectric, since in many materials the polarization cannot be switched before dielectric breakdown occurs. The key experiment to establish ferroelectricity is therefore the measurement of the ferroelectric hysteresis loop, i.e., the polarization $P$ as a function of the applied field $E$. Figure~\ref{fig:hysteresis_states} illustrates the idealized characteristic states I--V during domain switching for a z-cut crystal, which is the standard geometry for bulk LN.

For a hysteresis measurement, a crystal of thickness $d$ is contacted with electrodes on its $+z$ and $-z$ faces over the area $A$ [Fig.~\ref{fig:hysteresis_states}(a)]. The electrode material can significantly affect the measurement \cite{Wengler2003,miller1998periodically,Zahn2026}. For bulk LN, both in hysteresis measurements and in periodic poling, liquid electrodes (e.g., aqueous LiCl or NaCl solutions) are often used. The liquid contacts the surface homogeneously on a microscopic scale and enable a uniform charge exchange without local field spikes that would alter nucleation \cite{Koppitz2025,Wengler2003,Ratzenberger2024}. Metal electrodes, in contrast, may form a granular, rough interface with incomplete contact and strong local field enhancements. These can cause local switching, sparking, or dielectric breakdown well below the macroscopic (average) coercive field. Moreover, different electrode materials may result in different nucleation densities and charge-injection barriers \cite{miller1998periodically,Zahn2026}. This has to be taken into account when interpreting hysteresis loops.

A voltage $U$ applied to the electrodes generates the field $E = U/d$ in the crystal, while the current flowing to and from the electrodes is monitored. This can be realized, e.g., by measuring the voltage drop $U_R$ across a shunt resistor $R$ of well-known resistance, which is proportional to the current [Fig.~\ref{fig:hysteresis_states}(a)].

Single-domain crystals of LN are readily commercially available. Therefore, for LN, a hysteresis measurement usually starts with a single-domain crystal, i.e., the whole volume is polarized in one direction.  At $E = 0$, the crystal is thus in a single polarization state (state I), marked as I in the hysteresis loop in Fig.~\ref{fig:hysteresis_states}(c) and in the first sketch of Fig.~\ref{fig:hysteresis_states}(d). Note that $P$ and $E$ in hysteresis loops are taken along the direction of the applied field. Since the positive voltage is applied to the $+z$ face, the field is antiparallel to the polarization of the as-grown crystal, which therefore appears at $P=-P_S$ in the loop, although its polarization points along $+z$. When the voltage is increased, typically with a linear (triangular) ramp as shown in Fig.~\ref{fig:hysteresis_states}(b), initially no switching current flows and the polarization stays constant (apart from a small dielectric contribution). Once the field reaches a sufficient level, the first inverted domains nucleate and grow, and charge starts to flow, since the bound surface charge of each inverted region [Eq.~(\ref{P_s})] changes sign and has to be compensated. Note that the field at which switching sets in is \emph{not} the coercive field. The coercive field (state II) is defined as the field at which the average polarization of the whole contacted area vanishes, $P = 0$, i.e., when half of the area is switched and half of the total charge has flowed.

As the field increases further, more domains switch until the whole area $A$ is inverted. The current then stops, and the charge and the polarization saturate; apart from a small dielectric contribution, $P$ remains constant with further increasing field. When the field is removed, the polarization remains at $P = +P_S$ (state III).

Applying a field of opposite polarity reverses the process. Initially, the polarization stays constant until the first domains switch back. At $P = 0$, the reverse coercive field is reached, and half of the area is switched back (state IV). With further increasing field, the remaining domains switch back until the polarization saturates and the switching current returns to zero. After removal of the field (state V), the crystal is back in its initial configuration (state I).

An ideal hysteresis loop is symmetric and nearly rectangular, with vertical switching edges and horizontal saturation branches. Deviations, e.g., slanted edges or branches, may indicate delayed nucleation, defects, leakage currents, or incomplete switching, among others. Therefore a hysteresis measurements provides direct access to many in-situ physical process happening during switching.

Because the polarization of a switched region of area $A$ changes from $-P_S$ to $+P_S$, i.e., by $2P_S$, the total switched charge is \cite{Kumar2026}
\begin{equation}
\label{eq:charge}
Q = 2 \, P_S A .
\end{equation}
Hence, if the switched area $A$ is known, $P_S$ can be calculated from the integrated current (see next section). A switched charge lower than expected, a reduced polarization, or a loop that does not close can indicate a smaller switched area, charge losses by leakage currents, or a deviating stoichiometry that changes $P_S$. Combining hysteresis measurements with domain-imaging techniques, such as SHM \cite{Kirbus2024,Kiseleva2025,ruesing2019,Hegarty2022}, Raman imaging \cite{Rix2022,Reitzig2022}, polarization microscopy \cite{Ratzenberger2024,Merz1954}, selective etching \cite{Kumar2026}, or piezoresponse force microscopy (PFM) \cite{Soergel2011,Gruverman2019,Roeper2024}, provides in-depth insight into the ferroelectric behavior and the underlying domain physics of the crystal.

The coercive field, defined by $P(E_C) = 0$, depends on the stoichiometry and quality of the crystal. For the lithium niobate and lithium tantalate material family, it is well known that the coercive field strongly depends on defects, stoichiometry, and doping. The most common form of LN, the congruent composition, which contains a high density of intrinsic defects, has a high coercive field of about 21~kV/mm \cite{bocchini2025,Gopalan2007,Koppitz2025,Shur2006,Shur2016}, whereas values as low as 2~kV/mm have been reported for stoichiometric LN \cite{Kim2002}. For 5~mol\% MgO-doped LN, coercive fields as low as 4.5~kV/mm are typically reported \cite{bocchini2025,Gopalan2007,Koppitz2025,Ishizuki2003}. Moreover, in congruent LN it is well established that the coercive field for the first (forward) switching of the as-grown crystal differs significantly from that for reverse switching. This shifts the hysteresis loop along the field axis and is attributed to an internal field $E_\mathrm{int}$ caused by defect complexes that do not reorient during switching \cite{Gopalan2007,Gopalan1998}:
\begin{equation}
\label{eq:internal}
E_\mathrm{int} = \frac{E_{C,\mathrm{f}} - |E_{C,\mathrm{b}}|}{2} ,
\end{equation}
where $E_{C,\mathrm{f}}$ and $E_{C,\mathrm{b}}$ denote the forward and backward coercive fields, respectively. Consequently, switching a previously poled region back to its as-grown (virgin) orientation requires a smaller field. The internal field is close to zero for stoichiometric and 5~mol\% MgO-doped LN, but amounts to about 2~kV/mm for congruent LN and 5~kV/mm for congruent lithium tantalate \cite{Gopalan1996}. Comparing forward and backward coercive fields therefore provides insight into the defect structure of the crystal. As seen below, asymmetric hysteresis loops can also caused by the measurement dynamics \cite{Koppitz2025} and care must be taken when using Eq.~\ref{eq:internal} to interpret the internal electric field.

Crucially, such investigations and their interpretation are missing for (x-cut) TFLN, where discrepancies such as the high poling fields are not yet understood.

\subsection{Hysteresis loop measurement}

In practice, the polarization is obtained from the switched charge via Eq.~(\ref{P_s}). When a domain is inverted, the bound charges on its two ends change sign, and compensating charges have to flow through the external circuit. Since the polarization changes by $2P_S$, a total charge of $2P_S A$ flows for complete switching [Eq.~(\ref{eq:charge})]. Consequently, when the field approaches the coercive field, a switching current $I_P$ flows, as shown in Fig.~\ref{fig:hysteresis_states}(b). Integrating this current yields the \emph{change} of the polarization:
\begin{equation}
\label{eq:hysteresis}
\Delta P(t) = \frac{Q(t)}{A}= \frac{1}{A} \int_0^t I_P(t')\, \mathrm{d}t' .
\end{equation}
This is illustrated in Fig.~\ref{fig:hysteresis_states}(b), which shows the voltage pulse, the associated current, and the integrated charge. Note that this method only measures \emph{changes} of the polarization, not its absolute value. By convention, the measured loop is therefore shifted along the polarization axis such that it is symmetric, i.e., such that the saturation branches lie at $\pm P_S$.

In the hysteresis loop, the polarization is plotted as a function of the electric field $E$ instead of the time $t$. This is straightforward, since for each time $t_i$ the field $E(t_i) = U(t_i)/d$ is known from the applied voltage and the sample thickness or electrode gap $d$. The standard waveform for determining the coercive field is a triangular pulse, whose flanks have a constant ramp rate [Fig.~\ref{fig:hysteresis_states}(b)]:
\begin{equation}
\label{eq_rec}
\frac{\mathrm{d}E}{\mathrm{d}t} = \frac{1}{d}\frac{\mathrm{d}U}{\mathrm{d}t} = \mathrm{const.}
\end{equation}

A key challenge is that the measured current $I_{\text{total}}$ contains further contributions. Conventionally, two additional components are considered:
\begin{equation}
\label{eq:Itotal}
I_{\text{total}} = I_P + I_L + I_C ,
\end{equation}
where $I_P$ is the switching (poling) current, $I_L$ is the Ohmic leakage current, and $I_C$ is the capacitive current, which is also impacted by the permittivity of the sample. Here, high capacitive samples with strong dielectric responses show significant dielectric polarization contributions \cite{Scott1988}. Figure~\ref{fig:pund_method}(a) shows idealized versions of these components and their superposition.

The capacitive current follows from the capacitor equation,
\begin{equation}
\label{eq:IC}
I_C(t) = C\, \frac{\mathrm{d}U}{\mathrm{d}t} ,
\end{equation}
where $C$ is the capacitance of the sample and the circuit. For the linear ramp of Eq.~(\ref{eq_rec}), $I_C$ is constant because $\mathrm{d}U/\mathrm{d}t$ is constant and changes its sign when the ramp reverses [Fig.~\ref{fig:pund_method}(a)]. 

The leakage current is usually Ohmic,
\begin{equation}
\label{eq:IL}
I_L(t) = \frac{U(t)}{R_L} ,
\end{equation}
where $R_L$ is the resistance between the electrodes. It therefore grows linearly with the applied voltage [Fig.~\ref{fig:pund_method}(a)]. For thick bulk samples, both contributions are usually small and can often be neglected, since the voltage ramps are slow and the resistance is high. For TFLN, in contrast, fast voltage ramps are used and the electrodes are separated by only a few tens of micrometers, which can result in leakage along the surface. Since the switched charge is very small due to the small cross-section for TFLN, the capacitive and leakage currents can be comparable to or even exceed the switching current (see Sec.~\ref{sec:results}). They therefore have to be separated from the switching current.

% COMMENT: FIGURE 3a to be slightly changed:in this picture length of electrode (l_E) looks like it is labelling the +P_s arrow and the text of +P_s is floating away from its home arrow. so I was slightly confused until I saw the thinner arrow below the gold electrode. Please rearrange the texts next to the right arrows

\begin{figure}[t]
\centering
\includegraphics[width=\columnwidth]{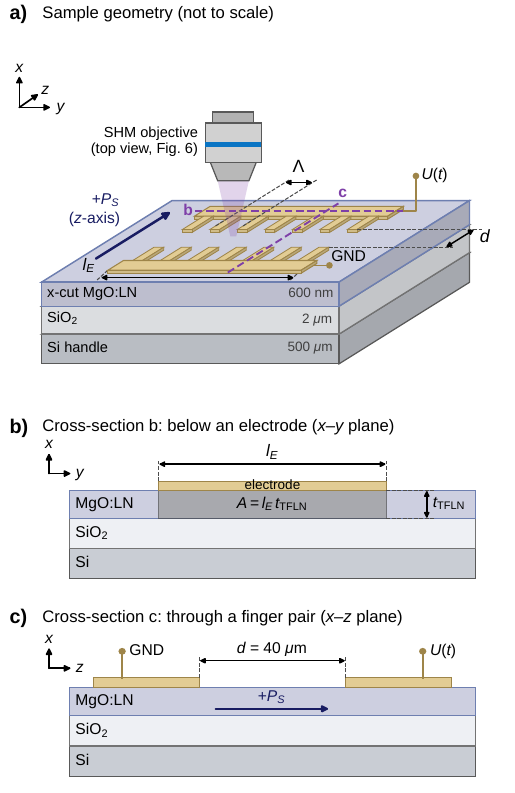}
\caption{(a) Schematic of the sample (not to scale): x-cut 5~mol\% MgO:TFLN (600~nm) on a 2-$\mu$m-thick SiO$_2$ layer on a Si handle with Cr/Au finger electrodes of length $l_E$ and period $\Lambda$, separated by the gap $d$ between the finger tips. The polar $z$-axis ($+P_S$) lies in the plane of the film. SHM images (Fig.~\ref{fig:hysteresis_SHG}) are recorded from the top. Dashed lines b and c mark the cross-sections shown in (b) and (c). (b) Cross-section below an electrode ($x$--$y$ plane) with the switched area $A = l_E\,t_\mathrm{TFLN}$. (c) Cross-section through a finger pair ($x$--$z$ plane), i.e., the side view used in Fig.~\ref{fig:hysteresis_SHG}.}
\label{fig:TFLN}
\end{figure}

The positive-up-negative-down (PUND) method separates the switching current from capacitive and leakage currents by a sequence of four voltage pulses \cite{Koppitz2025,Scott1988,Martin2017}. Table~\ref{tab:pund} summarizes the pulse sequence and the corresponding physical processes.

\begin{table}[t]
\centering
\caption{PUND pulse sequence: polarity of the four pulses (P, U, N, D), domain state before and after each pulse (starting from state I in Fig.~\ref{fig:hysteresis_states}; see also Fig.~\ref{fig:pund_method}(b)), and current components present during each pulse. }
\label{tab:pund}
\begin{tabular}{cccc}
\toprule
\textbf{Pulse} & \textbf{Polarity} & \textbf{\shortstack{Domain state\\ (before $\to$ after)}}  & \textbf{\shortstack{Current\\ components}} \\
\midrule
P & positive & $-P_S \to +P_S$ & $I_P + I_L + I_C$ \\
U & positive & $+P_S \to +P_S$ & $I_L + I_C$ \\
N & negative & $+P_S \to -P_S$ & $I_P + I_L + I_C$ \\
D & negative & $-P_S \to -P_S$ & $I_L + I_C$ \\
\bottomrule
\end{tabular}
\end{table}

The key principle is that during the second (U) and fourth (D) pulses, the sample is already in the final polarization state. Hence, no switching current flows, and only the capacitive, dielectric or leakage currents are measured, which are assumed to be independent of the domain orientation. Subtracting the currents of these reference pulses from those of the switching pulses (P and N) isolates the switching current:
\begin{align}
\label{eq:pund}
I_{\mathrm{sw}}^{+} &= I^{(\mathrm{P})} - I^{(\mathrm{U})} , \\
I_{\mathrm{sw}}^{-} &= I^{(\mathrm{N})} - I^{(\mathrm{D})} .
\end{align}
Figure~\ref{fig:pund_method}(b) illustrate the PUND method, showing the sequence of the voltage pulses and the corresponding current responses.

The coercive field is not a fixed material constant but depends on the measurement dynamics, in particular on the ramp rate \cite{Shur2014,Ganesamoorthy2005,Koppitz2025}. This is because domain nucleation and domain-wall motion are thermally activated, time-dependent processes, which depend on the electric field. At high ramp rates, less time is available for nucleation and growth. Switching then occurs at higher fields, where nucleation is more probable and domain walls move faster, both along and perpendicular to the polar axis. The domain-wall velocity typically follows a Merz-type exponential law as a function of the electric field \cite{miller1998periodically,Merz1954}. 

Correspondingly, the dependence of the coercive field on the ramp rate $\mathrm{d}U/\mathrm{d}t$ can be described empirically by
\begin{equation}
\label{eq:KAI}
E_C \left(\frac{\mathrm{d}E}{\mathrm{d}t} \right) = E_C^{\text{qs}}  + A_E \left(\frac{\mathrm{d}E}{\mathrm{d}t}\right)^\alpha ,
\end{equation}
where $E_C^{\text{qs}}$ is the coercive field in the quasi-static limit, and $A_E$ and $\alpha$ are material-dependent constants. For congruent LN, $\alpha$ typically ranges between 0.1 and 0.3 \cite{Shur2014,Orihara1994,Hashimoto1994}. Equation~(\ref{eq:KAI}) can be derived from a time-dependent formulation of the Kolmogorov--Avrami--Ishibashi (KAI) model \cite{Shur2014,Ishibashi1971}.

This dynamic effect is particularly important for TFLN, where the poling pulses are typically up to three orders of magnitude shorter, and the ramps correspondingly faster, than in bulk poling (Table~\ref{tab:ln_coercive_fields}).

\section{Experimental methods}
\label{sec:methods}

Samples were fabricated from x-cut, 5~mol\% MgO-doped TFLN wafers (NanoLN) with a film thickness of $t_\text{TFLN} = 600$~nm on a 2-$\mu$m-thick SiO$_2$ layer on a 500-$\mu$m-thick (100) Si substrate. Electrodes were patterned by electron-beam lithography and lift-off and consist of a 170-nm Cr layer and a 30-nm Au layer, following the layer structure of Ref.~\cite{Younesi2021}. The sample layout is shown in Fig.~\ref{fig:TFLN}. In contrast to the z-cut bulk geometry in Fig.~\ref{fig:hysteresis_states}, the polar $z$-axis lies in the plane of the film, and the electrodes are deposited on the x-face rather than on the $+z$ and $-z$ faces. Inverted domains therefore have to grow both along the surface and into the depth of the film. Moreover, charge has to be transferred into and out of the crystal to form and screen the charged head-to-head and tail-to-tail DWs that terminate each domain. In addition, periodic poling of TFLN is performed with finger electrodes of period $\Lambda$ (Fig.~\ref{fig:TFLN}). Here, each finger tip limits the contact surface for charge exchange. The switched area $A$ in Eq.~(\ref{eq:hysteresis}) is then given by the electrode length $l_E$ times the film thickness:
\begin{equation}
\label{eq:area}
A_\text{TFLN}=l_E \, t_\text{TFLN} .
\end{equation}
The gap $d$ between the opposing electrodes defines the (average) electric field. Note, that the field in the gap area is not homogenenous but is enhanced at finger tips \cite{Chen2024}. Even planar electrodes that are deposited on the planar x-cut surface have edges, where domains have to grow into the depth rather than perpendicular to the contact surface. This significantly differs from typical z-cut bulk poling geometries.

Since this work aims at TFLN under conditions used for periodic poling, two electrode geometries were investigated:
\begin{enumerate}
\item Finger electrodes with a total length of $l_E = 495$~$\mu$m, with fingers protruding by about 10~$\mu$m and a finger width of about 1.2~$\mu$m. The period is $\Lambda = 3$~$\mu$m, in line with typical periods in PPTFLN \cite{Bollmers2025,Zhao2020,Babel:25,Chang2016,Wang2018}, and the gap is $d = 40$~$\mu$m. The gap is measured as the closest distance between two opposing electrodes, i.e. finger tip to finger tip.
\item Planar (continuous) electrodes of the same length $l_E = 495$~$\mu$m and the same gap $d = 40$~$\mu$m. Here, nucleation can occur anywhere along the electrode edge and is not restricted to the finger tips. Here, the electric field is still not homogeneous when compared to z-cut bulk.
\end{enumerate}

Most experiments were performed with finger electrodes, since these are used for PPTFLN device fabrication (Table~\ref{tab:ln_coercive_fields}). The electrodes were contacted with tungsten needles in a probe station. The circuit consists of the high-voltage source providing the voltage $U$ and a shunt resistor $R$ in series with the sample to monitor the current, identical to the circuit in Fig.~\ref{fig:hysteresis_states}(a). To prevent electrical breakdown in air (dielectric strength of about 3~kV/mm), the samples were covered with an insulating fluid with a dielectric strength above 50~kV/mm.

The switched cross-sections in TFLN are small. While electrodes for devices are typically a few millimeters long \cite{Zhao2020,Zhao2020c,Wang2018,Bollmers2025}, we use $l_E = 495$~$\mu$m in this work. With the film thickness of 0.6~$\mu$m, this yields $A \approx 3.0\times10^{-10}$~m$^2$. With Eq.~(\ref{eq:charge}) and the literature value $P_S = 0.72$~C/m$^2$, the expected switched charge is $Q \approx 0.43$~nC. If this charge flows within about 1~ms (see Table~\ref{tab:ln_coercive_fields}), the average current is about 0.4~$\mu$A. Since switching occurs only during a fraction of the voltage ramp, peak currents of a few microamperes are expected. For comparison, switching a bulk crystal over 1~cm$^2$ involves a charge of about $1.4\times10^{-4}$~C, compared to $4.3\times10^{-10}$~C here, i.e., almost six orders of magnitude more.

Approximating the two electrodes as parallel wires at a distance of 40~$\mu$m with a radius on the order of the electrode thickness (0.2~$\mu$m) yields a capacitance of a few fF, which is negligible; since this estimate depends only logarithmically on the radius, it holds also for the actual flat electrode strips. At the slew rate of $\mathrm{d}U/\mathrm{d}t = 1600\text{V}/1~ms$, this yields a current on the order of 4~nA, which is much lower than the expected poling current. However, there are other sources of capacitance in the circuit. For example, the large contact pads of about 500~$\mu$m length and 300~$\mu$m are coupled vertically through SiO$_2$ layer and the conductive silicon wafer. This yields a capacitance on the order of 1~pF. This suggests a capacitive current on the order of $\mu$A similar to the poling current. This already justifies the requirement for the PUND method.

We therefore developed a setup that detects currents in the (sub-)microampere range on the high-voltage side at voltages above 1000~V. It consists of the following components:
\begin{enumerate}
\item Signal generation: a data-acquisition card (RedLab 1808X, 16-bit resolution, 200~kS/s) controlled by LabVIEW generates the voltage waveforms.
\item High-voltage amplification: a high-voltage amplifier (Trek 20/20C; gain 2000~V/V, maximum output $\pm20$~kV and $\pm20$~mA) amplifies the signal.
\item Current measurement: a precision shunt resistor of $R_{\text{shunt}} = 99.2$~k$\Omega$ in series with the sample converts the current into a voltage, which is recorded with an oscilloscope (Tektronix TDS 2024C, 8-bit vertical resolution, 200~MHz bandwidth). A current of 1~$\mu$A thus corresponds to about 0.1~V. The shunt resistor was calibrated by fitting its current--voltage characteristic between 0 and 10~V with an Ohmic resistor.
\end{enumerate}
The current is calculated from the voltage $U_R$ measured across the shunt:
\begin{equation}
\label{eq:current}
I = \frac{U_R}{R_{\text{shunt}}} .
\end{equation}

Unless stated otherwise, all hysteresis measurements used symmetric triangular pulses with a peak voltage of 1.6~kV, corresponding to a maximum field of 40~kV/mm for the 40-$\mu$m gap, rise and fall times of 1~ms each (total pulse duration 2~ms), and thus a ramp rate of $4\times10^{4}$~kV/(mm\,s). The delay between P and U or N and D pulses were typically more than 10~seconds to save and store the data.

Note that the electric field $E = U/d$ is only an estimate, since the field is enhanced near the finger tips and decreases with depth into the film \cite{Chen2024}. In the absence of a better estimate, we use $E \approx U/d$ in all following graphs.

For the ramp-rate study, ramp rates of $4\times10^{3}$, $4\times10^{4}$, $1\times10^{5}$, and $4\times10^{5}$~kV/(mm\,s) were used. The highest ramp rate was limited by the maximum slew rate of the combined limitations of the signal generator and high-voltage amplifier. The lowest ramp rate was limited by the noise of the current measurement, since the switching current $I_P$ [Fig.~\ref{fig:pund_method}(a)] decreases proportionally to the ramp rate.

Domain inversion was verified ex situ by SHM \cite{ruesing2019,Reitzig2021}, using a laser-scanning microscope (Zeiss LSM 980) with a femtosecond fiber laser at 920~nm (100-fs pulses, 80~MHz repetition rate, up to 1~W average power). The light was focused by a microscope objective (NA = 0.75), and the second-harmonic (SH) signal at 460~nm was collected in reflection. Domains of opposite polarization generate SH light with a phase difference of $\pi$. In TFLN, domains of either orientation therefore give the same signal as long as they extend through the full film thickness; only the DWs appear as dark lines, because the SH contributions from both sides of the wall interfere destructively within the focal volume. If a domain is not inverted through the full depth, i.e., if two domains of opposite orientation are stacked on top of each other, their SH contributions (partly) cancel, and the intensity with respect to a fully or un-inverted domain indicates the poling depth relative to the film thickness \cite{ruesing2019,Reitzig2021} [see the buried domain in Fig.~\ref{fig:hysteresis_SHG}(h)]. SHM thus confirms the poling depth and thereby the switched area $A$ used to calculate the spontaneous polarization.

The bulk reference measurement [Fig.~\ref{fig:hysteresis}(b)] was performed on a 200-$\mu$m-thick, single-domain z-cut 5~mol\% MgO:LN crystal (Yamaju Crystals, Japan). The crystal was contacted on both z faces with liquid electrodes, i.e., two tubes with an inner diameter of 2.5~mm (contact area 4.9~mm$^2$) filled with water saturated with NaCl as electrolyte. The electrodes were connected to a hysteresis measurement system (aixACCT TF Analyzer 2000) with a high-voltage measurement head and a high-voltage amplifier (Trek 10/10, Advanced Energy Inc.). A bipolar triangular waveform with an amplitude of 3~kV at 0.03~Hz was applied, corresponding to a ramp rate of 1.8~kV/(mm\,s).

\section{Results}
\label{sec:results}

\begin{figure*}[t]
\centering
\includegraphics[width=0.95\textwidth]{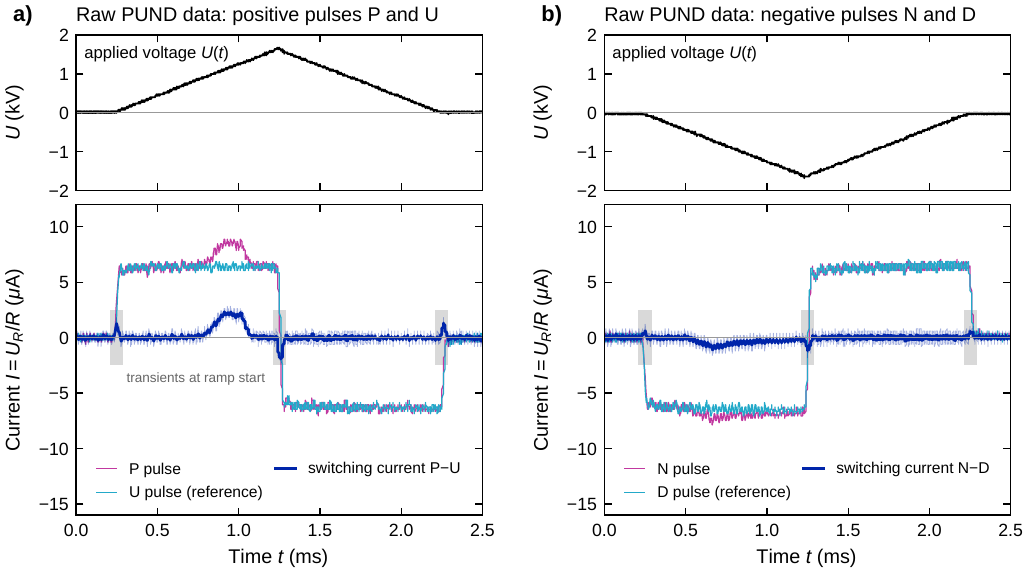}
\caption{Raw PUND data for finger electrodes at a ramp rate of $4\times10^{4}$~kV/(mm\,s). (a) Positive pulses: applied voltage $U(t)$ (top) and currents of the switching pulse P (magenta) and the reference pulse U (cyan), together with the switching current P$-$U (blue; light: raw data, dark: 15-$\mu$s moving average) (bottom). (b) The same for the negative pulses N and D. The currents were calculated from the voltage across the shunt resistor ($R = 99.2$~k$\Omega$). Gray boxes mark transients at the start, reversal, and end of the voltage ramps, which were excluded from the integration of the switching current.}
\label{fig:raw_pund}
\end{figure*}

\begin{figure*}[t]
\centering
\includegraphics[width=0.95\textwidth]{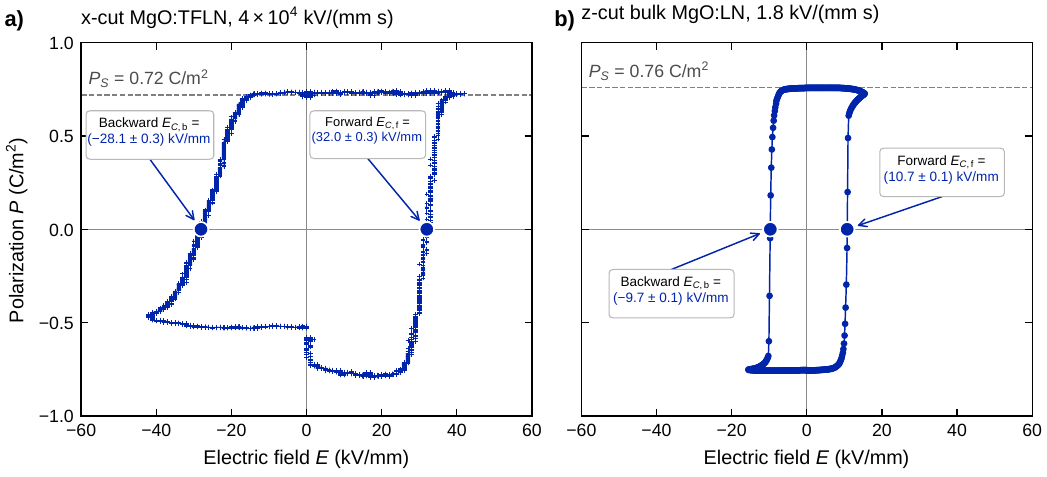}
\caption{(a) $P$--$E$ hysteresis loop of x-cut 5~mol\% MgO:TFLN with finger electrodes at a ramp rate of $4\times10^{4}$~kV/(mm\,s), obtained from the data in Fig.~\ref{fig:raw_pund} via Eq.~(\ref{eq:hysteresis}) and centered along the polarization axis. (b) For comparison, hysteresis loop of a 200-$\mu$m-thick, single-domain z-cut 5~mol\% MgO:LN crystal (Yamaju Crystals, Japan), measured with a bipolar triangular waveform at 0.03~Hz, i.e., a ramp rate of 1.8~kV/(mm\,s). Dots mark the forward and backward coercive fields $E_{C,\mathrm{f}}$ and $E_{C,\mathrm{b}}$, dashed lines the spontaneous polarization $P_S$. The errors of the coercive fields were determined from linear fits around the zero crossings and are dominated by the field resolution.}
\label{fig:hysteresis}
\end{figure*}

\subsection{Raw PUND data}

\begin{figure*}[t]
\centering
\includegraphics[width=0.95\textwidth]{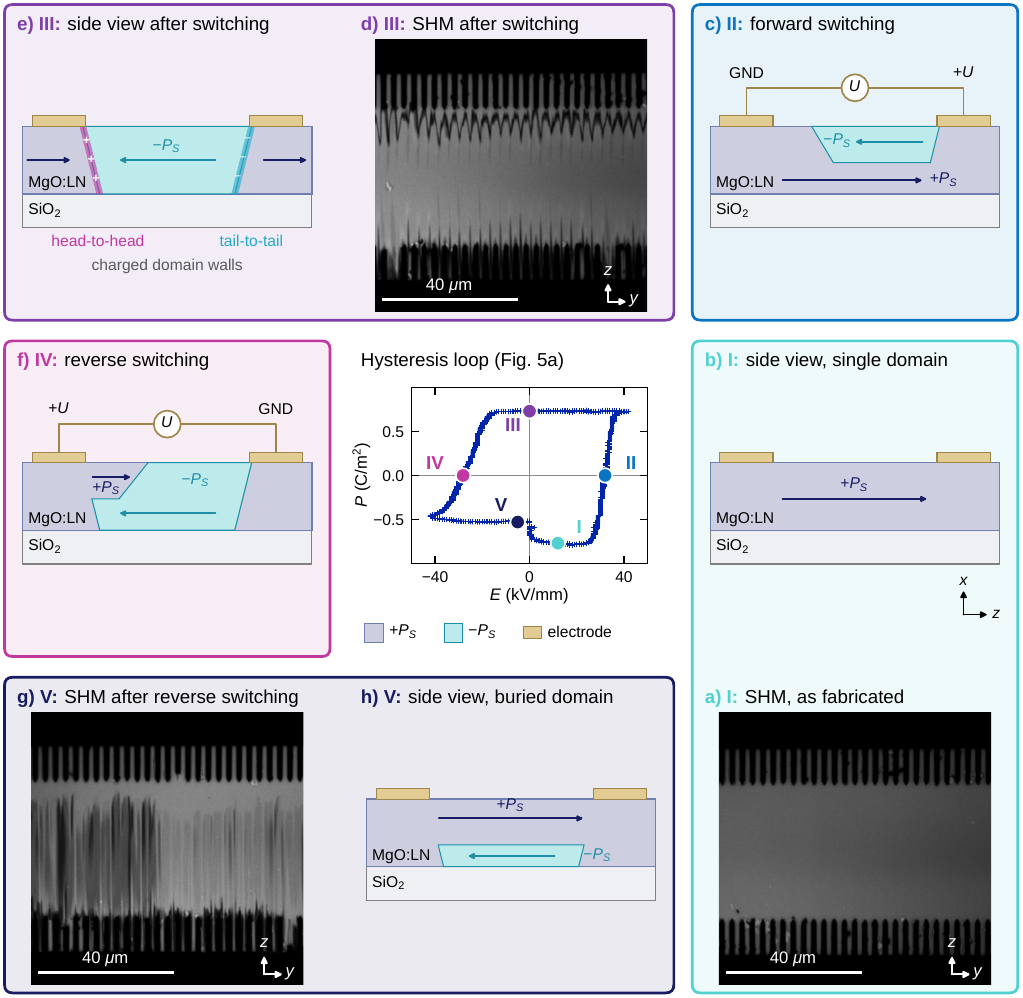}
\caption{Correlation of the measured hysteresis loop (center, data from Fig.~\ref{fig:hysteresis}(a)) with the domain configuration. The colored dots mark the states I--V in the loop; panels belonging to the same state are grouped in boxes of the same color. (a) SHM image of the as-fabricated sample and (b) side-view sketch of the initial single-domain state (state I); (c) sketch of forward switching near $E_{C,\mathrm{f}}$ (state II), with nucleation at the electrode on the $+z$ side and growth along the surface and into the depth; (d) SHM image after forward switching and (e) corresponding sketch with the charged head-to-head and tail-to-tail DWs (state III); (f) sketch of reverse switching near $E_{C,\mathrm{b}}$ (state IV), starting at the surface; (g) SHM image and (h) sketch after reverse switching (state V), showing a buried layer that remains inverted. SHM images show a top view ($y$--$z$ plane), sketches a side view ($x$--$z$ plane, not to scale). $\pm P_S$ in the sketches refers to the crystal $z$-axis, whereas the loop is plotted along the applied field (see Fig.~\ref{fig:hysteresis_states}). The domain shapes in the sketches are schematic: their depth and the inclination of the walls are only indicative. In any case, the walls that terminate the inverted domains inside the film are charged.}
\label{fig:hysteresis_SHG}
\end{figure*}

Figure~\ref{fig:raw_pund} shows raw PUND data for a typical sample with finger electrodes. The currents (left axis) were calculated from the voltage $U_R$ across the shunt resistor via $I = U_R/R$; the applied voltage is shown in the top panels. During the rising flank, all pulses show a constant capacitive current with a plateau of about 6.5~$\mu$A, which changes its sign during the falling flank, as expected from Eq.~(\ref{eq:IC}). For the ramp of $1.6\times10^{6}$~V/s, this corresponds to a capacitance of about 4~pF, which is presumably dominated by the probe needles, cables, and electrode pads. This fits well to the estimate discussed above of the order-of-magnitude via the coupled contact pads through the silicon. The plateaus of the U and D pulses show no noticeable slope, which indicates that Ohmic leakage is negligible for our samples. Clear switching peaks appear only during the P and N pulses, whereas the U and D pulses show only the capacitive contribution, in agreement with the expectations. Notably, the capacitive current exceeds the peak switching current (about 2--3~$\mu$A) by a factor of two to three, which highlights the necessity of the PUND method for TFLN. The difference currents (blue) isolate the pure switching current. Integrating the forward switching current yields a charge of $(0.43\pm0.01)$~nC consistent with the expected 0.43~nC.

Forward switching sets in at about 0.75~kV ($\approx 19$~kV/mm) and is completed at about 1.4~kV ($\approx 35$~kV/mm). Hence, already the onset of switching is about four times higher than the reported coercive field of bulk 5~mol\% MgO:LN \cite{Ishizuki2003}. The reverse switching current during the N pulse [Fig.~\ref{fig:raw_pund}(b)] is much weaker and broader. It sets in already at about $-0.5$~kV ($\approx -12.5$~kV/mm) and extends almost up to the maximum field of $-40$~kV/mm.

Between the positive (P, U) and negative (N, D) pulse pairs, typically 1--2~min passed for storing and checking the data and resetting the program. This waiting time can be important, since for bulk LN it is known that slow screening processes on a time scale of seconds can affect reverse switching immediately after forward poling; a waiting time on the order of minutes prevents this \cite{Koppitz2025}.

\subsection{Hysteresis loop and associated domain structure}

Using Eq.~(\ref{eq:hysteresis}), the accumulated switching charge was calculated for each time step and divided by the cross-sectional area to obtain the polarization. Each time step was assigned to the corresponding average field $E = U/d$, which yields the hysteresis loop shown in Fig.~\ref{fig:hysteresis}(a). The loop was centered along the polarization axis such that the original polarization and positive saturation polarizations are of equal magnitude. This is common practice \cite{Koppitz2025}, since only changes of the charge, not its absolute value, can be measured. Starting in the bottom-right corner, the loop shows the typical behavior: with increasing field, the polarization first remains constant. Switching sets in at about 20~kV/mm, although very weakly, and at the forward coercive field $E_{C,\mathrm{f}} = (32.0\pm0.3)$~kV/mm the polarization crosses zero. The coercive fields were determined by linear fits of $E(P)$ around the zero crossings; the uncertainty is dominated by the field resolution of 1~kV/mm and does not include the systematic uncertainty of the estimate $E = U/d$. This value agrees with the fields used for periodic poling of TFLN (Table~\ref{tab:ln_coercive_fields}), but it is substantially higher than the coercive fields of 4.5--10~kV/mm reported for bulk 5~mol\% MgO:LN \cite{bocchini2025,Hu2003,Koppitz2025,Ishizuki2003} and about three times the $(10.7\pm0.1)$~kV/mm that we obtain for a bulk z-cut 5~mol\% MgO:LN crystal at a ramp rate of 1.8~kV/(mm\,s) [Fig.~\ref{fig:hysteresis}(b)]. At about 35~kV/mm, the polarization saturates, and it remains constant when the field is removed, as expected for ideal ferroelectric switching. The resulting spontaneous polarization $P_S = (0.72\pm0.02)$~C/m$^2$ agrees remarkably well with values of 0.72--0.78~C/m$^2$ reported for congruent and MgO-doped LN \cite{Marchesano2015,Hu2003}. The agreement with the bulk value, together with the SHM results discussed below, confirms that the switched region extends over the full electrode length and film thickness. In contrast to TFLN, the bulk loop in Fig.~\ref{fig:hysteresis}(b) is nearly rectangular and symmetric ($E_{C,\mathrm{f}} = (10.7\pm0.1)$~kV/mm, $E_{C,\mathrm{b}} = (-9.7\pm0.1)$~kV/mm, $P_S \approx 0.76$~C/m$^2$) with steep switching edges. The rectangular shape reflects fast and homogeneous switching in the homogeneous field between the electrodes on the $+z$ and $-z$ faces.

Upon field reversal, reverse switching sets in at about $-12.5$~kV/mm, and the backward coercive field is reached at $E_{C,\mathrm{b}} = (-28.1\pm0.3)$~kV/mm. If the difference between forward and backward coercive fields is strictly interpreted as an internal field caused by non-switchable polar defect clusters, Eq.~(\ref{eq:internal}) yields $E_\mathrm{int} \approx 2$~kV/mm. This is comparable to the internal field typically reported for \emph{congruent} LN and significantly larger than that of 5~mol\% MgO-doped bulk LN, for which values of only 0 to 0.5~kV/mm are typically reported. The vanishing of the internal field is usually explained by Mg ions replacing the polar defect clusters and forming more mobile or non-polar defect complexes \cite{Koppitz2025,Gopalan2007,Prasath2021}. Based on the hysteresis loop alone, one might conclude that TFLN contains more defects causing internal fields, e.g., due to the ion-slicing process as suggested by some authors \cite{Boes2023} and subsequent annealing, which can lead to Li out-diffusion \cite{ZHANG2020b}. 

However, we are cautious with such an interpretation, since poling of TFLN differs from bulk poling in other respects as well. In TFLN, the domains do not terminate at the $z$-faces but at charged head-to-head and tail-to-tail walls within the film, and the electrodes are \emph{not} placed on the charged faces of the domains, but rather on the nonpolar surface. This can be expected to affect domain stability, nucleation, to create strong local field enhancements, and to hinder and cause incomplete screening after poling. An incompletely screened domain configuration also produces an internal (depolarizing) field that facilitates reverse switching. In this case, however, the internal field is caused by the incomplete screening of charged walls rather than by polar defects as in bulk crystals.

\begin{figure*}[t]
\centering
\includegraphics[width=0.95\textwidth]{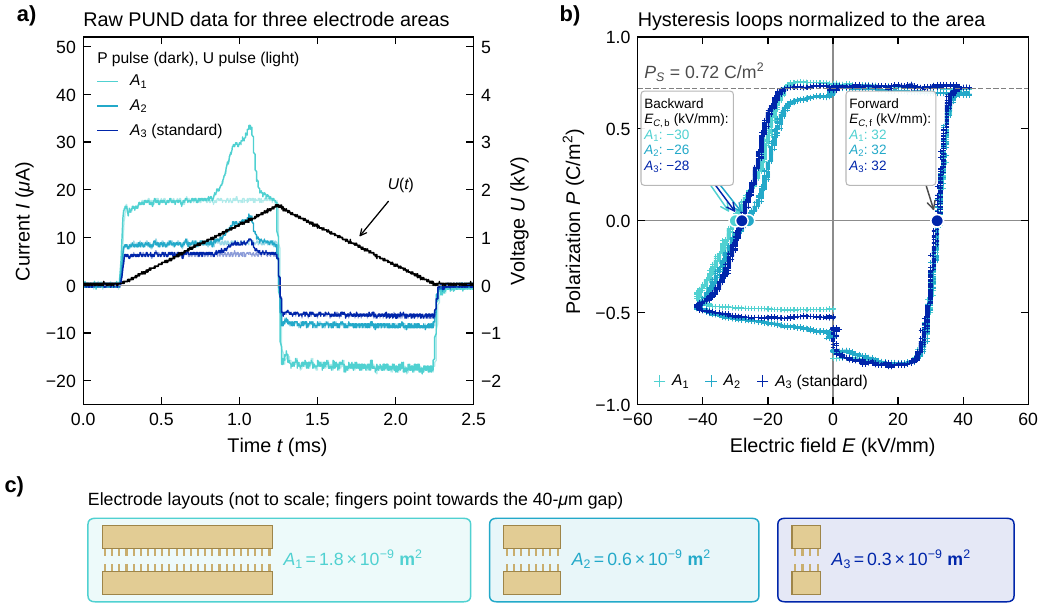}
\caption{Dependence on the electrode length. (a) Raw PUND data for the electrode areas $A_1 = 1.8\times10^{-9}$~m$^2$, $A_2 = 0.6\times10^{-9}$~m$^2$, and $A_3 = 0.3\times10^{-9}$~m$^2$ (standard electrode): currents of the P (dark) and U (light) pulses and applied voltage $U(t)$ (black, right axis); capacitive and switching currents scale with the electrode area. (b) Hysteresis loops normalized to the respective area, with the forward and backward coercive fields. (c) Electrode layouts (not to scale).}
\label{fig:electrode_length}
\end{figure*}

\begin{figure*}[t]
\centering
\includegraphics[width=0.95\textwidth]{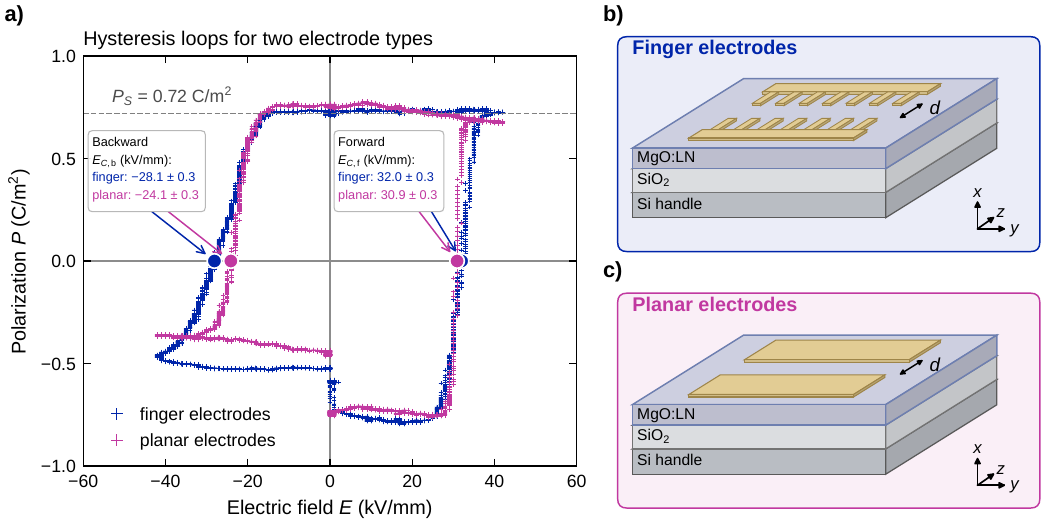}
\caption{Comparison of finger and planar electrodes. (a) Hysteresis loops for finger (blue) and planar (magenta) electrodes at $4\times10^{4}$~kV/(mm\,s), with the forward and backward coercive fields (finger: $E_{C,\mathrm{f}} = (32.0\pm0.3)$~kV/mm, $E_{C,\mathrm{b}} = (-28.1\pm0.3)$~kV/mm; planar: $E_{C,\mathrm{f}} = (30.9\pm0.3)$~kV/mm, $E_{C,\mathrm{b}} = (-24.1\pm0.3)$~kV/mm). (b), (c) Schematics of the finger and planar electrode geometries with gap $d$.}
\label{fig:planar_vs_finger}
\end{figure*}

Our data provide an argument for the latter interpretation: the apparent internal field is not a fixed quantity of the film but depends on the electrode geometry. For planar electrodes (Fig.~\ref{fig:planar_vs_finger}), we obtain $E_\mathrm{int} \approx 3.4$~kV/mm, and it also varies with the electrode length (between about 1 and 3~kV/mm, Fig.~\ref{fig:electrode_length}). A defect-related internal field, in contrast, should be a material property independent of the electrode geometry. This suggests that the loop asymmetry in TFLN is dominated by extrinsic effects, such as incomplete screening and hindered charge exchange or the measurement methodology, rather than by intrinsic polar defect complexes. Nevertheless, further studies of domain stability, which has been reported to be reduced under certain conditions in TFLN \cite{Sabatti2025,Shao2016,Zhang2020,Jiao2020}, and of the influence of defects and stoichiometry are required to clarify this point.

When the negative field is increased further, the polarization saturates at about $P = -0.5$~C/m$^2$ and remains at this value after the field is removed. The loop is thus not closed: the reverse-switched charge corresponds to only about 85\% of the forward-switched charge, i.e., about 15\% of the switched volume remains inverted. In additional experiments on other samples, we increased the reverse switching field to up to 100~kV/mm, but we never achieved a closed loop before electrical breakdown and sparking occurred. This indicates that some domains are very hard to switch back.

To confirm the poling result, to determine the switched area $A$, and to gain insight into the domain structure, we performed SHM imaging before poling [Fig.~\ref{fig:hysteresis_SHG}(a)], after forward poling [Fig.~\ref{fig:hysteresis_SHG}(d)], and after reverse switching [Fig.~\ref{fig:hysteresis_SHG}(g)]. Figure~\ref{fig:hysteresis_SHG}(a) (bottom right) shows an SHM image of the region between the finger electrodes before poling. The electrode fingers at the top and bottom of the image appear black, since the metal does not generate a second-harmonic signal. The LN film in between shows a strong, homogeneous signal without any DW contrast, which is the typical SHM signature of single-domain TFLN \cite{ruesing2019,Reitzig2021}. After forward poling [Fig.~\ref{fig:hysteresis_SHG}(d)], most of the gap still shows a homogeneous SH signal, while DW signatures appear near the finger electrodes at the top and bottom of the image. The central region between the fingers shows the same intensity as before poling and no substructure. This clearly indicates a fully inverted domain that extends through the entire film thickness, which justifies calculating the polarization with the full cross-sectional area, as discussed above. 

After reverse switching [Fig.~\ref{fig:hysteresis_SHG}(g)], however, the image is no longer homogeneous: dark lines and gray regions remain in the center of the gap. This indicates domains of opposite orientation stacked on top of each other, i.e., the surface layer has switched back, while a buried layer has remained inverted. This agrees well with the open hysteresis loop. 

In detail, examining the contrast of the dark areas relative to the fully reverse-switched area yields an SHM signal intensity reduction of 20-50\% depending on the evaluated area. According to Rüsing et al. \cite{ruesing2019} this can be used to estimate the fraction of domain that has the opposing domain orientation. The data from \cite{ruesing2019} indicates approximately 70\% to 90\% of the film which has a different domain orientation. This fits well with the about 85\% estimated from the hysteresis loop.

To interpret these images, we first recall the established picture of domain inversion in LN and other ferroelectrics, in which the inversion process is divided into five main stages \cite{Shur2014,Shur2016,Shur2006,Shur2021}: (1) nucleation of inverted domains, (2) fast forward growth of needle-like domains along the polar axis ($z$-axis), (3) sideways growth perpendicular to the polar axis, (4) coalescence of neighboring domains, and (5) spontaneous back-switching if sufficient screening or other stabilization, e.g., by defect pinning, is not achieved before the field is removed.

For bulk z-cut LN, it is known that domain nucleation occurs at local field enhancements, e.g., at electrode edges \cite{hirohashi2006characterization,Shao2016}, at surface scratches, or at inhomogeneities in the crystal, such as local defects. From these nucleation sites, spike domains start to grow. In LN (MgO-doped or undoped), domain growth usually starts from the $+z$ face \cite{Shur2006,mackwitz2016,Zhao2020b}. This is attributed to different charge-injection barriers and to conductive DWs\cite{Werner2017,Schroeder2012}. In recent years, in particular the role of conductive DWs in the screening process during domain growth has been noted \cite{Kosobokov2024,Sturman2022,Podivilov2024,Bollmers2024}. Here, the initial spike domain, which in the initial phase has no contact with the opposing electrode, can only achieve screening via the conductive DW, which is in contact with the electrode during nucleation. As only positively charged (and hence negatively screened) DWs \cite{Ratzenberger2024,Beccard2022,Beccard2023,Singh2022,Kirbus2019} in MgO:LN are highly conductive, this facilitates the growth of initial domain spikes from the +z side only, where the spike has a slight angle with respect to the polar axis facilitating domain conductivity \cite{Ratzenberger2024,Singh2022,Kirbus2019}. Consistent with bulk crystals, it is well established that in x-cut TFLN the domains start to grow from the electrode on the $+z$ side, to which the positive voltage is applied \cite{Zhao2020b,Zhao2020c,Zhao2019}.

Applying this knowledge to x-cut TFLN with surface electrodes suggests the following sequence during forward poling: Nucleation starts at the finger tips of the electrode on the $+z$ side, where the field is highest. The initial spike domain is located at the surface and is in contact with the finger [Fig.~\ref{fig:hysteresis_SHG}(c)]. It grows rapidly along the $z$-axis close to the surface and subsequently expands into the depth ($x$-direction) and laterally ($y$-direction). When poling is complete, the inverted domain extends through the whole film thickness and terminates in charged head-to-head and tail-to-tail walls below or close to the electrodes [Fig.~\ref{fig:hysteresis_SHG}(e)]. This may be incompletely screened leading to the observed internal field. Note that the sketches show the bound charges; If fully screened, the associated mobile screening charges of opposite sign render the walls effectively neutral. The injection and exchange of these screening charges is what is measured as switching current in the external circuit.

When a reverse voltage is applied, nucleation of domains with the original orientation is expected to start at the opposite electrode, which is now at positive potential. Again, a spike domain nucleates in contact with the electrode, propagates along the surface, and subsequently expands laterally and into the depth. Because reverse switching starts at the surface, a layer of the inverted orientation remains buried beneath the reverse-switched surface layer [Fig.~\ref{fig:hysteresis_SHG}(g) and (h)]. Since this buried domain and its charged walls are no longer in direct contact with the electrodes, but only connected to them via the nominally insulating LN, the exchange of screening charges is drastically slowed down. Consequently, the reverse switching current is much lower and spread over a much wider field range than the forward switching current [Fig.~\ref{fig:raw_pund}(b)], and disproportionately high fields is required to switch the buried domains. This explains why a layer of buried domains remains which cannot be switched back even at very high fields [Fig.~\ref{fig:hysteresis_SHG}(g)]. It also highlights an important aspect of domain inversion: a field above the (local) coercive field is necessary, but not sufficient; the exchange of screening charges is required as well.

\subsection{Electrode-length dependence}

To separate geometry-related from material-related effects, we varied geometric and pulse parameters. The simplest variation is the electrode length (Fig.~\ref{fig:electrode_length}). A longer electrode should increase the switched charge proportionally to the cross-section $A$, while the hysteresis loops normalized to $A$ should be identical. Indeed, the capacitive and switching currents measured via the shunt voltage $U_R$ increase with the electrode length and contact pad size [Fig.~\ref{fig:electrode_length}(a)], and the normalized loops are nearly identical, in particular for forward switching [Fig.~\ref{fig:electrode_length}(b)]. For reverse switching, an additional, weak and non-monotonic dependence is observed, with backward coercive fields between $-26$ and $-30$~kV/mm. A voltage drop along the electrodes can be ruled out as the cause: for currents of a few microamperes and an electrode resistance of at most a few hundred ohms, the voltage drop is below 1~mV and thus negligible compared with the applied 1.6~kV. A possible explanation is leakage associated with the larger domain structures, e.g., via conductive DWs, since the effect appears only during reverse switching. Importantly, no such dependence is observed for forward switching, which indicates that poling recipes developed for short electrodes can be transferred directly to longer electrodes.

\subsection{Planar vs.\ finger electrodes}

So far, all measurements were performed with finger electrodes, as typical for periodic poling. To assess the role of the fingers, we fabricated planar electrodes of similar length and identical gap (40~$\mu$m) and measured them with the same procedure. Planar electrodes lack the local field enhancement at the finger tips, but they provide a continuous electrode edge at which nucleation can occur anywhere; both aspects may affect the hysteresis loop. Note, even for planar electrode edges, the electric field will be not as homogeneous as for bulk, z-cut poling.

Figure~\ref{fig:planar_vs_finger} compares the hysteresis loops for both geometries. At first glance, the loops are very similar. Hence, the overall switching behavior is not dramatically affected by the use of fingers, and to first order hysteresis measurements on finger electrodes can be compared directly to those on planar electrodes, which are closer to a standard geometry for bulk hysteresis measurements.

On closer inspection, however, subtle differences emerge that are consistent with the picture developed above. For planar electrodes, forward switching sets in at a higher average field but then proceeds more abruptly, i.e., within a narrower field range, so that the coercive field is about 1~kV/mm lower than for finger electrodes [$(30.9\pm0.3)$ vs.\ $(32.0\pm0.3)$~kV/mm]. The earlier onset for finger electrodes can be explained by the field enhancement at the finger tips, which triggers nucleation already at a lower average field in the gap. Once switching has started, however, it proceeds also faster for planar electrodes, which we attribute to the larger number of possible nucleation sites and the larger electrode area available for charge exchange.

Reverse switching shows a similar picture: for planar electrodes, it sets in earlier and proceeds faster, resulting in a smaller backward coercive field [$E_{C,\mathrm{b}} = (-24.1\pm0.3)$~kV/mm vs.\ $(-28.1\pm0.3)$~kV/mm]. At the same time, the remanent polarization after reverse switching is smaller in magnitude (about $-0.38$~C/m$^2$ vs.\ $-0.47$~C/m$^2$), i.e., a larger fraction of the domains remains inverted. This suggests that nucleation and the initial charge exchange at the surface are easier for planar electrodes, whereas the electrically isolated domains at the bottom of the film are harder to switch back. A possible reason is the missing field enhancement at the finger tips, which promotes local charge injection or conductivity. In contrast, for planar electrodes the switching close to the surface happens faster and more thoroughly due to a larger number of spike domains leading to more buried domains out of electrical contact.

During forward switching, both geometries reach nearly the same spontaneous polarization, which confirms that the switched cross-sections are identical.

In summary, the macroscopic switching dynamics observed in the hysteresis measurements are similar for finger and planar electrodes. This has implications for the small periods that are increasingly used, e.g., for counter-propagating phase-matching schemes \cite{Sabatti2025,Zhao2020c}. A higher density of fingers might bring the poling behavior closer to that of planar electrodes, i.e., switching within a narrower field range, which would require a more precise control of the applied voltage than for longer periods.

\begin{figure*}[t]
\centering
\includegraphics[width=0.95\textwidth]{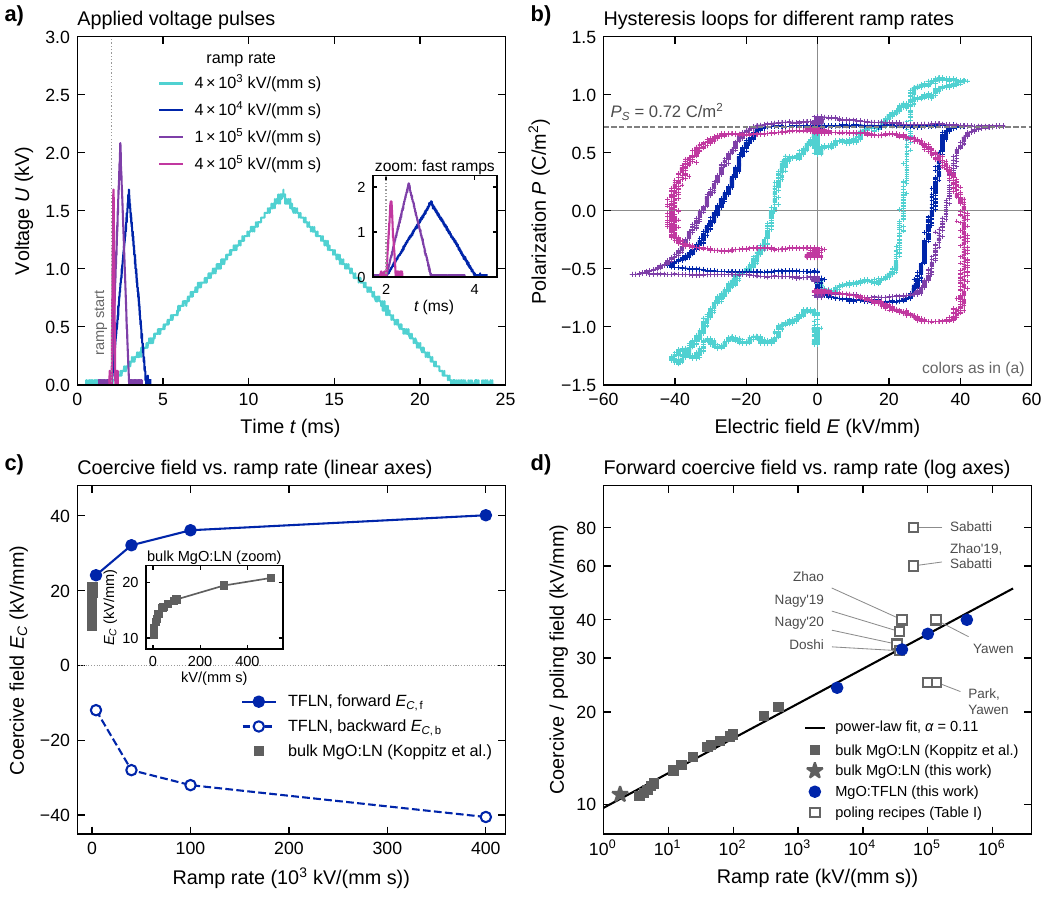}
\caption{Ramp-rate dependence of the coercive field (finger electrodes). (a) Applied triangular voltage pulses for ramp rates of $4\times10^{3}$ to $4\times10^{5}$~kV/(mm\,s), shifted to a common ramp start (inset: the three fastest ramps). (b) Corresponding hysteresis loops. (c) Forward (filled) and backward (open) coercive fields of TFLN as a function of the ramp rate on linear axes, together with the forward coercive fields of bulk z-cut 5~mol\% MgO:LN from Koppitz \emph{et al.}~\cite{Koppitz2025} (gray; inset: zoom-in). (d) Forward coercive fields on double-logarithmic axes: TFLN (blue), bulk MgO:LN from Ref.~\cite{Koppitz2025} (gray squares) and from Fig.~\ref{fig:hysteresis}(b) (star), and power-law fit $E_C = A_E\,(\mathrm{d}E/\mathrm{d}t)^{\alpha}$ (black line), which appears as a straight line on these axes. Open squares: peak fields of the periodic-poling recipes for MgO:TFLN from Table~\ref{tab:ln_coercive_fields}, labeled by first author: Park~\cite{Park2022}, Zhao~\cite{Zhao2020,Zhao2020b,Zhao2020c,ruesing2019}, Zhao'19~\cite{Zhao2019}, Nagy'19~\cite{Nagy2019}, Nagy'20~\cite{Nagy2020}, Doshi~\cite{Doshi2024}, Sabatti~\cite{Sabatti2025}, Yawen~\cite{Yamen2026}.}
\label{fig:ramp_rate}
\end{figure*}
% FIGURE TODO (Fig. 9): English labels ("Ramprate" -> "Ramp rate", "Thinfilm-Lithiumniobat" -> "thin-film LN",
%  "Bulk-Lithiumniobat" -> "bulk LN", "Allometric fit curve" -> "power-law fit"); decimal points; panel titles
%  "Ramprate over Coercive Field E_C 1/2" -> e.g. "Coercive field vs. ramp rate (linear/log scale)";
%  unit "kV/(s*mm)" -> "kV/(mm s)"; remove the text-box "Koppitz et al." arrow or make it consistent with other labels.

\subsection{Ramp-rate dependence}

As discussed in Sec.~\ref{sec:theory}, the coercive field depends on the ramp rate [Eq.~(\ref{eq:KAI})]. The ramp rates used for periodic poling of TFLN are two to three orders of magnitude higher, and the poling times correspondingly shorter, than in bulk z-cut experiments (Table~\ref{tab:ln_coercive_fields}). To be comparable to typical PPTFLN poling, we therefore used a relatively high ramp rate of $4\times10^{4}$~kV/(mm\,s) so far, and indeed observed the high fields also in the hysteresis loops. To test whether the coercive field of TFLN follows the same ramp-rate dependence as that of bulk LN \cite{Shur2014,Koppitz2025}, we performed measurements at $4\times10^{3}$, $4\times10^{4}$, $1\times10^{5}$, and $4\times10^{5}$~kV/(mm\,s). Even the lowest of these ramp rates is still about one order of magnitude higher than the highest ramp rate in the bulk study by Koppitz \emph{et al.} \cite{Koppitz2025}.

Figure~\ref{fig:ramp_rate}(a) shows the applied voltage pulses: from the slowest ($4\times10^{3}$~kV/(mm\,s), turquoise) via the standard pulse used above (blue) to the two faster pulses (purple and magenta). Note that the peak voltages of the faster pulses deviate from the nominal 1.6~kV (about 2.1~kV and 1.7~kV, respectively), which was due to discretizations in the pulse generation. However, the ramp rate or switching behavior itself is not influenced by the maximum voltage. Figure~\ref{fig:ramp_rate}(b) shows the corresponding hysteresis loops. The loop at $4\times10^{4}$~kV/(mm\,s) (blue) is identical to Fig.~\ref{fig:hysteresis}(a). In contrast, the faster ramps result in clearly larger coercive fields, and the slowest ramp (turquoise) results in a drastically reduced coercive field for both forward and backward switching. The asymmetry and noise of the loops show that the measurements at $4\times10^{3}$ and $4\times10^{5}$~kV/(mm\,s) are at the limits of our experimental setup. At the lowest ramp rate, the switching current approaches the noise floor, and integrating small offsets over the 25-ms-long pulse leads to a drift, which is visible as an apparent polarization exceeding $P_S$ and an open loop. At the highest ramp rate, the voltage ramp is sampled less steps leading to a reduced accuracy and distorted shape. Nevertheless, these measurements allow us to extract the trend of the coercive field as function of the ramp rate.

Figure~\ref{fig:ramp_rate}(c) shows the forward and backward coercive fields as a function of the ramp rate on linear axes, together with the data of Koppitz \emph{et al.} for bulk z-cut 5~mol\% MgO:LN \cite{Koppitz2025} (inset: zoom-in on the bulk data). Already this graph shows that our data reproduces the trend known from bulk samples \cite{Koppitz2025,Shur2014} for both forward and backward switching. This becomes even more striking on a double-logarithmic scale [Fig.~\ref{fig:ramp_rate}(d)]: the bulk data of Koppitz \emph{et al.} and our TFLN data lie approximately on a single straight line, as expected for a power law $E_C \propto (\mathrm{d}E/\mathrm{d}t)^{\alpha}$. A fit of the forward coercive fields of both data sets with Eq.~(\ref{eq:KAI}), neglecting the quasi-static term ($E_C^\mathrm{qs} = 0$), yields $\alpha = 0.11\pm0.01$. This value lies within the range of 0.1--0.3 reported for congruent bulk LN \cite{Shur2014,Orihara1994,Hashimoto1994}; to our knowledge, no values have been reported for MgO-doped bulk LN or TFLN so far.

Figure~\ref{fig:ramp_rate}(d) additionally shows the peak fields of those periodic-poling recipes for MgO:TFLN in Table~\ref{tab:ln_coercive_fields} for which a ramp rate is reported. Almost all of them lie on or above the fitted curve. This is expected, since poling recipes are designed to complete domain inversion reliably, and in a short time and therefore apply fields somewhat above the coercive field for the respective ramp rate. Recipes with very narrow, sidewall-electrode gaps \cite{Sabatti2025} or films bonded to silicon photonics \cite{Zhao2019} lie furthest above the curve, consistent with the geometric contributions and the uncertainty of the estimate $E = U/d$ discussed above. Only two recipes lie below the curve \cite{Park2022,Yamen2026}; Yamen et al. performed poling of ridge structures and under non-standard conditions, i.e., at elevated temperature, which reduce the coercive field \cite{Yamen2026}. In contrast Park, applied a short initial pulse (0.3 ms), but applied a long tail ($>20$~ms), which might result in the growth of domains. Additionally, their film was relatively thin (250~nm compared to 500~nm or more in Table~1), while the period was already less then 1.5~$\mu$m which might required altered poling conditions to avoid merging. In contrast, most works used periods in the order of 3~$\mu$m. The fitted curve thus provides a practical lower bound for the field required for poling typical TFLN at a given ramp rate.

The fact that a single power law describes both the bulk data, measured on z-cut crystals with liquid electrodes, and our TFLN data, measured with coplanar finger electrodes, over more than four orders of magnitude of ramp rate strongly suggests that the high coercive fields observed in TFLN are mainly a consequence of the high ramp rates. Two caveats should be noted. First, there are no data between $5\times10^{2}$ and $4\times10^{3}$~kV/(mm\,s), where the two data sets would overlap. Second, the geometry-related contributions discussed above (about 1~kV/mm due to the finger electrodes and the uncertainty of the estimate $E = U/d$) are small compared with the ramp-rate effect, but not negligible. 

These results are consistent with the work of Liu \emph{et al.} \cite{Liu2026}, who periodically poled undoped x-cut TFLN with long pulses of up to 500~ms and required a maximum field of only 26.6~kV/mm (Table~\ref{tab:ln_coercive_fields}), i.e., only about 5~kV/mm above the coercive field of bulk congruent LN. Similarly,  Zhang \emph{et al.} \cite{Zhang2020} achieved poling with 28.6~kV/mm with up to 1000 pulses of 50~ms. Overall, our results show that the high coercive fields observed in TFLN are not primarily an intrinsic material effect, as previously suggested in the literature, but largely a consequence of the poling dynamics.

\section{Discussion}
\label{sec:discussion}

Our measurements reveal several key similarities and differences between x-cut TFLN and z-cut bulk LN:

\begin{itemize}

\item \emph{Spontaneous polarization:} The measured $P_S = 0.72$~C/m$^2$ of MgO-doped TFLN matches bulk values of 0.72 to 0.78~C/m$^2$ \cite{bocchini2025,Koppitz2025}. This indicates that the ferroelectric order is preserved in the thin film and gives no indication of a reduced polarization, e.g., due to Li out-diffusion.

\item \emph{Coercive fields:} The coercive field of $(32.0\pm0.3)$~kV/mm under our standard conditions is about seven times the quasi-static bulk value of about 4.5~kV/mm \cite{Koppitz2025,Ishizuki2003,bocchini2025}. However, this comparison is misleading, since the ramp rates differ by orders of magnitude and TFLN is poled with coplanar (finger) surface electrodes instead of electrodes on the $\pm z$ faces.

\item \emph{Hysteresis loop shape:} Bulk hysteresis loops are nearly rectangular owing to the homogeneous field between the electrodes and the planar electrodes with large areas for nucleation and charge exchange on the $\pm z$ faces. TFLN loops are more rounded and slanted owing to the inhomogeneous field of the surface electrodes, especially for the case of finger electrodes, which leads to an earlier onset of nucleation but a slower charge exchange.

\item \emph{Asymmetry and incomplete reverse switching:} TFLN shows an apparent internal field of a few kV/mm and an open hysteresis loop, with about 15\% of the switched volume remaining inverted after reverse switching. Since the apparent internal field depends on the electrode geometry, we attribute it mainly to incomplete screening and hindered charge exchange rather than to polar defects.

\end{itemize}

Our systematic investigation identifies three factors contributing to the high coercive fields observed in TFLN, which we discuss in the order of their importance.

The most significant factor is the use of fast poling pulses. Typical TFLN poling uses ramp rates of $10^{4}$--$10^{5}$~kV/(mm\,s), up to three orders of magnitude faster than bulk poling (Table~\ref{tab:ln_coercive_fields}). Figure~\ref{fig:ramp_rate} shows that the coercive fields of TFLN and bulk MgO:LN follow a common power law as a function of the ramp rate, so that extrapolating the TFLN data to bulk ramp rates yields coercive fields close to the bulk values. This strongly indicates that the high fields are not intrinsic to TFLN but a consequence of the switching dynamics. The physical origin of this ramp-rate dependence is the finite time required for nucleation and domain-wall motion: at high ramp rates, the system has to be driven harder to complete switching within the available time. This is a universal feature of ferroelectrics and not specific to TFLN.

\begin{itemize}
\item \emph{Finger electrodes:} Finger electrodes, which are essential for periodic poling, restrict nucleation and charge injection to discrete finger tips. The comparison in Fig.~\ref{fig:planar_vs_finger} shows that finger electrodes increase the forward coercive field by about 1~kV/mm and broaden the switching transition. This effect is mainly geometric: finger electrodes create a highly inhomogeneous field that is concentrated at the tips. Nucleation therefore occurs primarily at the tips, and the resulting spike domains have to grow and coalesce laterally, which requires higher fields and more time.

\item \emph{Surface-electrode geometry and charge exchange:} The coplanar surface electrodes used for x-cut TFLN differ fundamentally from the electrodes on the $\pm z$ faces of z-cut bulk crystals. With surface electrodes, the electric field decays with depth into the film \cite{Chen2024}, and the charge exchange with the full film thickness is impeded. The SHM images before and after poling suggest that nucleation starts at the surface and that regions near the substrate interface are inverted last. During reverse switching, these regions become electrically isolated once the surface layer has switched back, which prevents further charge exchange, e.g., via charged and conductive DWs. The buried domains are therefore very difficult to switch back, which explains the incomplete reverse switching and the open hysteresis loops. Crucially, this effect does not occur during forward poling, which is the process relevant for periodic poling. For forward poling, the surface-electrode geometry likely contributes to the high fields via the decay of the field with depth, which requires a higher average field to invert the full film thickness. Our experiments cannot quantify this contribution separately; field simulations or measurements with modified electrode geometries (e.g., electrodes on etched sidewalls \cite{Sabatti2025,Franken2026}) could clarify it. Simulations suggest that the field decay is only a few percent translating to an effect of one kV/mm or less \cite{Chen2024,Zhao2019}.
\end{itemize}

\subsubsection*{Implications for periodic poling} 
Our results suggest that slower poling pulses could substantially reduce the voltage required for periodic poling of TFLN. It should be kept in mind, however, that slower switching also leaves more time for sideways domain growth, so that the optimum pulse shape likely represents a trade-off between required field and domain fidelity, which remains to be investigated.

\subsubsection*{Implications for ferroelectric and DW electronics}
The results translate directly to TFLN-based DW devices, which use similar coplanar electrodes and charged walls. First, the coercive field is not a material constant: it changes by almost a factor of two across the investigated ramp rates and follows a power law. For DW memories and memristors, this defines the trade-off between write speed and write voltage and provides a simple handle on the coercive field, whose control is a key challenge also for emergent ferroelectric thin films \cite{Mikolajick2021}. Second, the apparent internal field (imprint) depends on the electrode geometry, which indicates that it is governed by screening and charge exchange at electrodes and charged walls rather than by bulk defects; electrode design and interface engineering are thus levers to control imprint and retention. Third, reverse switching in the coplanar geometry leaves buried, electrically isolated domains. For in-plane DW devices, the erase operation may thus leave residual charged walls, which could contribute to cycle-to-cycle variability and limited endurance; sidewall electrodes as used for periodic-poling in optics \cite{Sabatti2025,Franken2026} or longer erase pulses may mitigate this. Fourth, partial switching, which sets the duty cycle in periodic poling, is also the basis of multilevel DW memristors \cite{McConville2020}; the combination of PUND and SHM provides a calibrated measure of the switched volume, i.e., of the internal state variable of such devices. Finally, TFLN combines a single-crystalline film with well-established bulk properties. It is therefore a model system to separate intrinsic material properties from extrinsic effects of geometry and dynamics, which is more challenging in polycrystalline ferroelectrics.

\section{Conclusion}
\label{sec:conclusion}

We have presented the first systematic hysteresis measurements of x-cut MgO-doped thin-film lithium niobate. Using a custom measurement setup and the PUND method, we measured a spontaneous polarization of $(0.72\pm0.02)$~C/m$^2$, consistent with bulk values, and a forward coercive field of $(32.0\pm0.3)$~kV/mm for a ramp rate of $4\times10^{4}$~kV/(mm\,s).

By systematically varying the electrode geometry and the ramp rate, we have shown that the high coercive fields observed in periodic poling of TFLN are not intrinsic material properties. They result mainly from (i) the use of poling pulses up to three orders of magnitude faster than in bulk poling and, to a lesser extent, from (ii) finger electrodes that limit nucleation and charge transfer and (iii) the coplanar surface-electrode geometry, which impedes charge exchange with the full film thickness and leads to incomplete reverse switching.

The ramp-rate dependence is particularly significant: the coercive fields of TFLN and bulk MgO:LN follow a common power law, and extrapolating the TFLN data to bulk ramp rates yields coercive fields close to the bulk values. This indicates that the high fields required for TFLN poling are a consequence of the switching dynamics rather than of thin-film-specific material properties.

Future work could investigate the effect of annealing or of pre-pulses, which have been reported to improve periodic poling results \cite{Younesi2021,Nagy2020,Doshi2024}, although the underlying mechanisms are not fully understood. In this regard, hysteresis measurements could help to separate the effects of defects, internal fields, or electrode geometry. Measurements at intermediate ramp rates and with rectangular pulses of different durations would further connect the TFLN and bulk data. Moreover, in-situ SHM during poling would directly visualize the switching dynamics and could validate the proposed charge-exchange mechanisms. In-situ SHM has been demonstrated for bulk samples \cite{Kirbus2019,Kirbus2024,Kiseleva2025} but require substantial improvement in time resolution over current SHM techniques for TFLN. Here, novel approaches like ultrafast widefield SHM could be applied in the future \cite{Vittadello2021}.

Beyond periodic poling, the results are relevant for TFLN-based ferroelectric and DW electronics: the ramp-rate dependence of the coercive field, the geometry-dependent imprint, and the incomplete reverse switching of buried domains directly affect the write and erase operation, retention, and endurance of DW memories, memristors, and neuromorphic devices in the same coplanar geometry. Extending the present approach to repeated cycling, minor loops and partial switching, temperature-dependent measurements, and combined measurements of switching charge and DW conduction is therefore a natural next step. 

More than half a century after its first systematic growth, LN thus remains an emergent material, and understanding its switching is a common key to its photonic and electronic applications.

\begin{acknowledgments}
The DFG (FOR5044 -- Project No.~426703838) is gratefully acknowledged for financial support. This work was supported by the German Federal Ministry of Research, Technology and Space (BmFTR) under the joint project Quantenrepeater.Net (QR.N). N. Vishnukumar acknowledges support via the Max Planck School of Photonics. Parts of this manuscript were drafted and improved with Claude Opus 5.5.
\end{acknowledgments}

% Bibliography generated from references.bib with BibTeX and included directly (arXiv does not run BibTeX)

\end{document}